\documentclass[useAMS,usenatbib]{mn2e}

\usepackage[T1]{fontenc}
 \usepackage{pslatex}
\usepackage{amsmath}
\usepackage{amsfonts}
\usepackage{color}
\usepackage{url}
\usepackage{graphicx}
\usepackage{subfigure}
\usepackage{amssymb}
\usepackage{soul}
\usepackage{booktabs}
\usepackage{array}
\usepackage[utf8]{inputenc}
\inputencoding{latin1}
\inputencoding{utf8}

{\newif\ifnotend
\notendtrue
\def\veclist{ABCDEFGHIJKLMNOPQRSTUVWXYZabcdefghijklmnopqrstuvwxyz.}
\def\top#1#2.{#1}
\def\tail#1#2.{#2.}
\loop\expandafter\xdef\csname v\expandafter\top\veclist\endcsname%
{{\noexpand\bf\expandafter\top\veclist}}
\edef\veclist{\expandafter\tail\veclist}
\if\veclist.\notendfalse\fi\ifnotend\repeat}
\def\e{{\rm e}}
\def\i{{\rm i}}

\def\E{{\cal E}}

\mathchardef\mhyphen="2D

\title[Instability of Magnetised Force-Free Jets]{External Confinement and Surface Modes in Magnetised Force-Free Jets}
\author[Sobacchi \& Lyubarsky]{E. Sobacchi$^{1,2}$\thanks{E-mail: sobacchi@post.bgu.ac.il} \& Y. E. Lyubarsky$^1$\\
$^1$ Physics Department, Ben-Gurion University, P.O.B. 653, Beer-Sheva 84105, Israel \\
$^2$ Department of Natural Sciences, The Open University of Israel, 1 University Road, P.O.B. 808, Raanana 4353701, Israel
}
\begin{document}

\date{}

\def\p{\partial}
\def\E{\textbf{E}}
\def\B{\textbf{B}}
\def\v{\textbf{v}}
\def\j{\textbf{j}}
\def\s{\textbf{s}}
\def\e{\textbf{e}}

\newcommand{\di}{\mathrm{d}}
\newcommand{\bfx}{\mathbf{x}}
\newcommand{\bfe}{\mathbf{e}}
\newcommand{\vlos}{\mathrm{v}_{\rm los}}
\newcommand{\Tspin}{T_{\rm s}}
\newcommand{\Tb}{T_{\rm b}}
\newcommand{\degree}{\ensuremath{^\circ}}
\newcommand{\Th}{T_{\rm h}}
\newcommand{\Tc}{T_{\rm c}}
\newcommand{\bfr}{\mathbf{r}}
\newcommand{\bfv}{\mathbf{v}}
\newcommand{\bfu}{\mathbf{u}}
\newcommand{\pc}{\,{\rm pc}}
\newcommand{\kpc}{\,{\rm kpc}}
\newcommand{\Myr}{\,{\rm Myr}}
\newcommand{\Gyr}{\,{\rm Gyr}}
\newcommand{\kms}{\,{\rm km\, s^{-1}}}
\newcommand{\de}[2]{\frac{\partial #1}{\partial {#2}}}
\newcommand{\cs}{c_{\rm s}}
\newcommand{\rb}{r_{\rm b}}
\newcommand{\rqu}{r_{\rm q}}
\newcommand{\bfOmega}{\pmb{\Omega}}
\newcommand{\bfOmegap}{\pmb{\Omega}_{\rm p}}
\newcommand{\bfXi}{\boldsymbol{\Xi}}

\maketitle

\begin{abstract}
In the paradigm of magnetic launching of astrophysical jets, instabilities in the MHD flow are a good candidate to convert the Poynting flux into the kinetic energy of the plasma. If the magnetised plasma fills the almost entire space, the jet is unstable to helical perturbations of its body. However, the growth rate of these modes is suppressed when the poloidal component of the magnetic field has a vanishing gradient, which may be the actual case for a realistic configuration.
Here we show that, if the magnetised plasma is confined into a limited region by the pressure of some external medium, the velocity shear at the contact surface excites unstable modes which can affect a significant fraction of the jet's body. We find that when the Lorentz factor of the jet is $\Gamma\sim10$ ($\Gamma\sim 100$), these perturbations typically develop after propagating along the jet for tens (hundreds) of jet's radii. Surface modes may therefore play an important role in converting the energy of the jet from the Poynting flux to the kinetic energy of the plasma, particularly in AGN. The scaling of the dispersion relation with (i) the angular velocity of the field lines and (ii) the sound speed in the confining gas is discussed.
\end{abstract}



\section{Introduction}
\label{sec:introduction}

Astrophysical jets are ubiquitous in a wide variety of events and they reach relativistic speeds in microquasars (e.g. \citealt{MirabelRodriguez1999}), Active Galactic Nuclei (AGN; e.g. \citealt{UrryPadovani1995}) and Gamma Ray Bursts (GRBs; e.g. \citealt{Piran2004}). One of the most promising explanations for the launching mechanism is energy extraction from a rotating, magnetised source (e.g. \citealt{Blandford1976, Lovelace1976, BlandfordZnajek1977}): the plasma slides along the magnetic field lines anchored to the central object, and is accelerated by magnetic tension.

In this scenario, one of the key points is the fate of the electromagnetic fields at large distances from the source. On one hand, in a steady, axisymmetric, ideal MHD flow, both analytical and numerical works have shown that the plasma can be accelerated up to approximate equipartition (i.e. to a magnetisation $\sigma\approx 1$), but achieving further acceleration is generally difficult (e.g. \citealt{Komissarov2007, Komissarov2009, Lyubarski2009, Lyubarski2010, Lyubarski2011, Tchekhovskoy2008, Tchekhovskoy2009, Tchekhovskoy2010}). However, observations of both GRBs \citep{ZhangKobayashi2005, Mimica2009b, Mimica2009a, MimicaAloy2010, Narayan2011} and AGN \citep{Ghisellini2010, Tavecchio2011} suggest the jet's magnetisation to be well below unity in the point where most of the radiation is emitted.

Hence, in the context of the magnetic launching paradigm, it is important to identify some mechanism to convert the jet's energy from the Poynting flux into the kinetic energy of the plasma. Instabilities in the MHD flow are indeed a good candidate to destroy the regular structure of the jet and cause the release into the plasma of the energy stored in the electromagnetic fields (e.g. \citealt{Lyubarski1992, Eichler1993, Spruit1997, Begelman1998, Giannios2006}).

In the simplest configuration, the magnetised plasma fills the entire space and pressure equilibrium is guaranteed by the gradients of the electromagnetic fields. In this case, modes growing with time are generally concentrated at the core of the jet (in the following we are referring to these as ``core modes''). However, these perturbations become stable if the poloidal component of the magnetic field has a vanishing gradient (e.g. \citealt{IstominPariev1996, Lyubarski1999, Mizuno2012, Sobacchi2017}).\footnote{Note that in this case the hoop stress of the toroidal magnetic field is balanced by the electric field, and not by the gradient of the poloidal magnetic field. This is indeed possible for a relativistic MHD flow.} This might indeed be the case for realistic configurations of the electromagnetic fields (see the discussion in \citealt{Narayan2009}).

Hence, it is important to identify other potential sources of instability. If the jet is confined to some limited region by the pressure of the surrounding gas (e.g. \citealt{Lynden-Bell1996, Lynden-Bell2003}), the contact surface may become unstable due to the velocity shear between the relativistic plasma and the confining medium (in the following we are referring to these modes as ``surface modes''). In the context of magnetised jets, these modes have been investigated analytically by \citet{Hardee2007}, who considered a non-rotating jet with a constant longitudinal magnetic field. \citet{Narayan2009} relaxed these assumptions, but assumed a rigid wall at the jet's boundary (or, equivalently, that the confining gas is cold).

Here we present a systematic study of surface modes in the case of relativistically rotating, force-free jets. We use an idealised setup with a sharp transition between the relativistic, magnetised plasma and the confining gas. We find that the combination of (i) the sound speed of the confining gas and (ii) the angular velocity of the field lines plays a critical role in determining the dispersion relation. We also show that these modes can perturb a significant fraction of the jet's body.

The paper is organised as follows. In Section \ref{sec:equations} we present the fundamental equations describing the steady-state solution and the evolution of the perturbations. In Section \ref{sec:results} we present the dispersion relation of surface modes, and its dependence on the relevant physical parameters. Finally, in Section \ref{sec:conclusions} we summarise our conclusions.

\section{Fundamental equations}
\label{sec:equations}

The fundamental equations governing astrophysical jets are the Maxwell's equations
\begin{align}
\label{eq:maxwell_1}
\nabla\times\E & = -\frac{1}{c}\frac{\p\B}{\p t} & \nabla\cdot\B & =0 \\
\label{eq:maxwell_2}
\nabla\cdot\E & =4\pi\rho& \nabla\times\B & =\frac{4\pi}{c}\j +\frac{1}{c}\frac{\p\E}{\p t}\;,
\end{align}
where $\rho$ and $\j$ are the charge and current densities. In the ideal MHD approximation, these are coupled with the condition of infinite conductivity
\begin{equation}
\label{eq:cond}
\E+\frac{\v}{c}\times\B=0 \;,
\end{equation}
where $\v$ is the velocity of the flow. When the energy of the flow is almost entirely carried in the form of Poynting flux, we have a force-free configuration and Euler's fluid equation reduces to
\begin{equation}
\label{eq:euler_forcefree}
\rho\E + \frac{\j}{c}\times\B=0\;.
\end{equation}

We assume that the confining gas is non-magnetised, and has a politropic equation of state with adiabatic index $\Gamma=5/3$ (or $\Gamma=~4/3$ for a relativistically hot gas),
\begin{equation}
\label{eq:adiabatic_EOS}
p_{\rm gas} = K\rho_{\rm gas}^{\Gamma}\;,
\end{equation}
where $p_{\rm gas}$ ($\rho_{\rm gas}$) is the pressure (proper mass density) of the gas and $K$ is a constant. The governing equation is the Euler's equation
\begin{equation}
\label{eq:euler_gas}
\frac{\partial\v_{\rm gas}}{\partial{\rm t}} + \left(\v_{\rm gas}\cdot\nabla\right)\v_{\rm gas} = -\frac{\nabla p_{\rm gas}}{w_{\rm gas}}\;,
\end{equation}
where $\v_{\rm gas}$ is the velocity and $w_{\rm gas}$ is the enthalpy of the gas, with
\begin{equation}
w_{\rm gas}=\rho_{\rm gas}+\frac{\Gamma}{\Gamma-1}\frac{p_{\rm gas}}{c^2}\;,
\end{equation}
and the continuity equation
\begin{equation}
\label{eq:continuity}
\frac{\partial\rho_{\rm gas}}{\partial{\rm t}} + \nabla\cdot\left(\rho_{\rm gas}\v_{\rm gas}\right) = 0\;.
\end{equation}
In the following we take a uniform $\rho_{\rm gas}$, and assume that the confining gas is at rest (i.e. $\v_{\rm gas}=0$) in the steady-state; this justifies neglecting the bulk Lorentz factor of the flow and the time derivative of the gas pressure in Eq. \eqref{eq:euler_gas} and \eqref{eq:continuity}.

\subsection{Unperturbed solution}
\label{sec:background}

From Eq. \eqref{eq:maxwell_1}-\eqref{eq:euler_forcefree} one can derive the equation for the steady-state equilibrium configuration of a cylindrical jet. It is
\begin{equation}
\label{eq:equilibrium}
r^2\frac{\text{d}B_{\rm z}^2}{\text{d}r}+\frac{\text{d}}{\text{d}r}\left(r^2B_\phi^2\right)-\frac{\text{d}}{\text{d}r}\left(\frac{\Omega^2r^4B_{\rm z}^2}{c^2}\right) =  0 \;,
\end{equation}
where 
\begin{equation}
\label{eq:defomega}
\Omega \equiv \frac{1}{r}\left(v_\phi - \frac{B_\phi}{B_{\rm z}} v_{\rm z}\right)
\end{equation}
is the angular velocity of the field lines (with this definition $E_{\rm r}~=~\Omega rB_{\rm z}/c$). In this paper we are neglecting any dependence of $\Omega$ on $r$.

Unless differently specified (as in Section \ref{sec:Bprofile}), through this paper we are using a flat profile for the longitudinal magnetic field, i.e. $B_{\rm z}$ does not depend on $r$. In this case, the solution of Eq. \eqref{eq:equilibrium} is simply
\begin{equation}
\label{eq:Bphi_flat}
B_\phi = -\frac{\Omega r}{c}B_{\rm z}\;.
\end{equation}
It is well known that a force-free, cylindrical equilibrium where $B_{\rm z}$ does not depend on $r$ is stable to the core modes \citep{IstominPariev1996, Lyubarski1999, Mizuno2012, Sobacchi2017}; hence, this is the ideal case to isolate the effect of the surface modes, which are the main focus of this paper.

To study the effect of the external confinement, we take a force-free configuration of the electromagnetic fields (that is, a solution of Eq. \ref{eq:equilibrium}) at radii $r<R$; in the following we are often referring to $R$ as the radius of the jet. We assume a sharp transition to occur at $r=R$, where a confining medium is balancing the inner magnetic pressure. Note that a current (charge) sheet at the jet's boundary is required to account for the jump of the magnetic (electric) field. The equilibrium condition is
\begin{equation}
\label{eq:balance}
p_{\rm mag}=p_{\rm gas}\;,
\end{equation}
where $p_{\rm mag}$ ($p_{\rm gas}$) is the pressure of the electromagnetic fields (of the confining gas) at $r=R$. Since the pressure is a Lorentz scalar, we need to take
\begin{equation}
\label{eq:p_mag}
p_{\rm mag}=\frac{B'^2}{8\pi}=\frac{B^2-E^2}{8\pi}\;,
\end{equation}
where $B'$ is the magnetic field in the frame of the magnetised plasma (in this frame one has $E'=0$ due to the condition of infinite conductivity). Finally, note that for the jet model defined by Eq. \eqref{eq:Bphi_flat} one finds $p_{\rm mag}=B_{\rm z}^2/8\pi$.

After the Poynting-dominated plasma expands beyond the light cylinder, its velocity approaches a pure drift motion, i.e. ${\bf v}/c={\bf E}\times {\bf B}/B^2$ (e.g. \citealt{Tchekhovskoy2009}). The corresponding Lorentz factor is
\begin{equation}
\label{eq:gammabeta}
\gamma\beta = \frac{E}{\sqrt{B^2-E^2}} = \frac{\Omega r}{c}\;,
\end{equation}
where the last equality is valid for a flat $B_{\rm z}$ profile. Since $\Omega R/c\gg 1$, the magnetised plasma moves with a relativistic velocity, while the confining medium is at rest. Hence, one would expect the surface at $r=R$ to become unstable to surface modes.

\subsection{Linearised equations for the perturbations}
\label{sec:lin}

We investigate the stability of cylindrical equilibria considering perturbations on the electromagnetic fields of the form $\exp\left[\i\left(kz + m\phi - \omega t\right)\right]$. To find the equation for the evolution of these perturbation we follow the framework developed by \citet{Solovev1967}, and generalised to the relativistic, force-free case by \citet{Lyubarski1999}. This calculation is more easily performed in the reference frame of the perturbation (i.e., a frame moving with velocity $u\equiv\omega/kc$), where the perturbed electromagnetic fields do not depend on time. The electromagnetic fields in the new frame are
\begin{equation}
\label{eq:field_trans}
E_{\rm r}^*=\frac{E_{\rm r}-uB_\phi}{\sqrt{1-u^2}} \qquad B_{\rm z}^*=B_{\rm z} \qquad B_\phi^*=\frac{B_\phi-uE_{\rm r}}{\sqrt{1-u^2}}\;.
\end{equation}
In this frame, Eq. \eqref{eq:maxwell_1}-\eqref{eq:euler_forcefree} are satisfied by
\begin{align}
\label{eq:Etrans}
\E^*=&-\nabla\phi \\
\label{eq:Btrans}
\B^*=& \frac{1}{\eta}\left(\nabla\psi\times\s + I\s\right) \;,
\end{align}
where $\eta\equiv m^2+k^2r^2\left(1-u^2\right)$ and $\s\equiv m\e_{\rm z} + kr\sqrt{1-u^2}\e_\phi$. It is possible to show that satisfying Eq. \eqref{eq:maxwell_1}-\eqref{eq:euler_forcefree} also requires $\nabla\psi\cdot\nabla\phi=0$ and $\nabla\psi\cdot\nabla I=0$. Hence, it is possible to introduce a new variable $\xi$ such that $\phi=\phi\left(\xi\right)$, $\psi=\psi\left(\xi\right)$, $I=I\left(\xi\right)$. Since $\s\cdot\nabla\psi=0$ due to helical symmetry, the magnetic field $\B^*$ is always perpendicular to $\nabla\psi=\psi'\nabla\xi$, and therefore $\xi$ remains constant on the magnetic surfaces.

In the steady-state solution magnetic surfaces are cylinders, in which case we are taking $\xi=r$. One can then study the evolution of the perturbations considering small deviations of $\xi$ from the unperturbed solution,
\begin{equation}
\xi = r - f\left(r\right)\exp\left[\i\left(kz + m\phi - \omega t\right)\right]\;.
\end{equation}
The time evolution of these perturbations is described by a second order linear differential equation, namely
\begin{equation}
\label{eq:main}
\frac{\text{d}}{\text{d}r}\left[G\frac{\text{d}f}{\text{d}r}\right]=Df\;.
\end{equation}
The functions $G$ and $D$ can be expressed in terms of the unperturbed solution as
\begin{align}
\label{eq:G}
G & \equiv \frac{r^3}{1-u^2}\left(\frac{a}{\beta}-b\right) \\
\label{eq:D}
D & \equiv k^2\left[\left(1-u^2+\frac{m^2-1}{k^2r^2}\right)G - d - \frac{1}{\beta}\frac{\text{d}}{\text{d}r}\left(r^4b\right)\right] \;,
\end{align}
where
\begin{align}
a & \equiv \left[\left(1-muV-u^2\right)B_{\rm z}-\frac{m}{kr}B_\phi\right]^2 \\
b & \equiv \left[VB_{\rm z}+\frac{u}{kr}B_\phi\right]^2 \\
d & \equiv \frac{2r^3}{\beta^2} \left[ \left(muVB_{\rm z}+\frac{m}{kr}B_\phi\right)^2  -  \left(1-u^2\right)^2B_{\rm z}^2\right] \;.
\end{align}
Here we have defined $V\equiv\Omega/kc$ and $u\equiv\omega/kc$.

The initial conditions can be found requiring that $f$ is not singular at $r=0$. Expanding Eq. \eqref{eq:main} it can be easily seen (e.g. \citealt{Sobacchi2017}) that when $m=1$ the proper condition is $f'\left(0\right)=0$, together with an arbitrary normalisation, for example $f\left(0\right)=1$; when $m\neq 1$ we can instead take $f\left(0\right)=0$ and $f'\left(0\right)=1$. Hence, we have a standard Cauchy problem, and one can integrate Eq. \eqref{eq:main} to find the solution for any $r$. Numerical integration is effectively started from a positive $r_{\rm init}\ll R$, and Taylor expansion of the solution is used to find the proper initial conditions.

In general, the solution of Eq. \eqref{eq:main} needs to satisfy two homogeneous boundary conditions, namely (i) $f\left(0\right)=0$ or $f'\left(0\right)=0$, depending on the value of $m$; (ii) the pressure equilibrium at $r=R$, which is a linear equation involving $f\left(R\right)$ and $f'\left(R\right)$, see Eq. \eqref{eq:PB} below. The value of $\omega$ is therefore automatically determined by these requirements, and can be found with the standard shooting method for eigenvalue problems (e.g. \citealt{Press2002}).

\begin{figure*}{\vspace{3mm}} 
\centering
\includegraphics[width=0.49\textwidth]{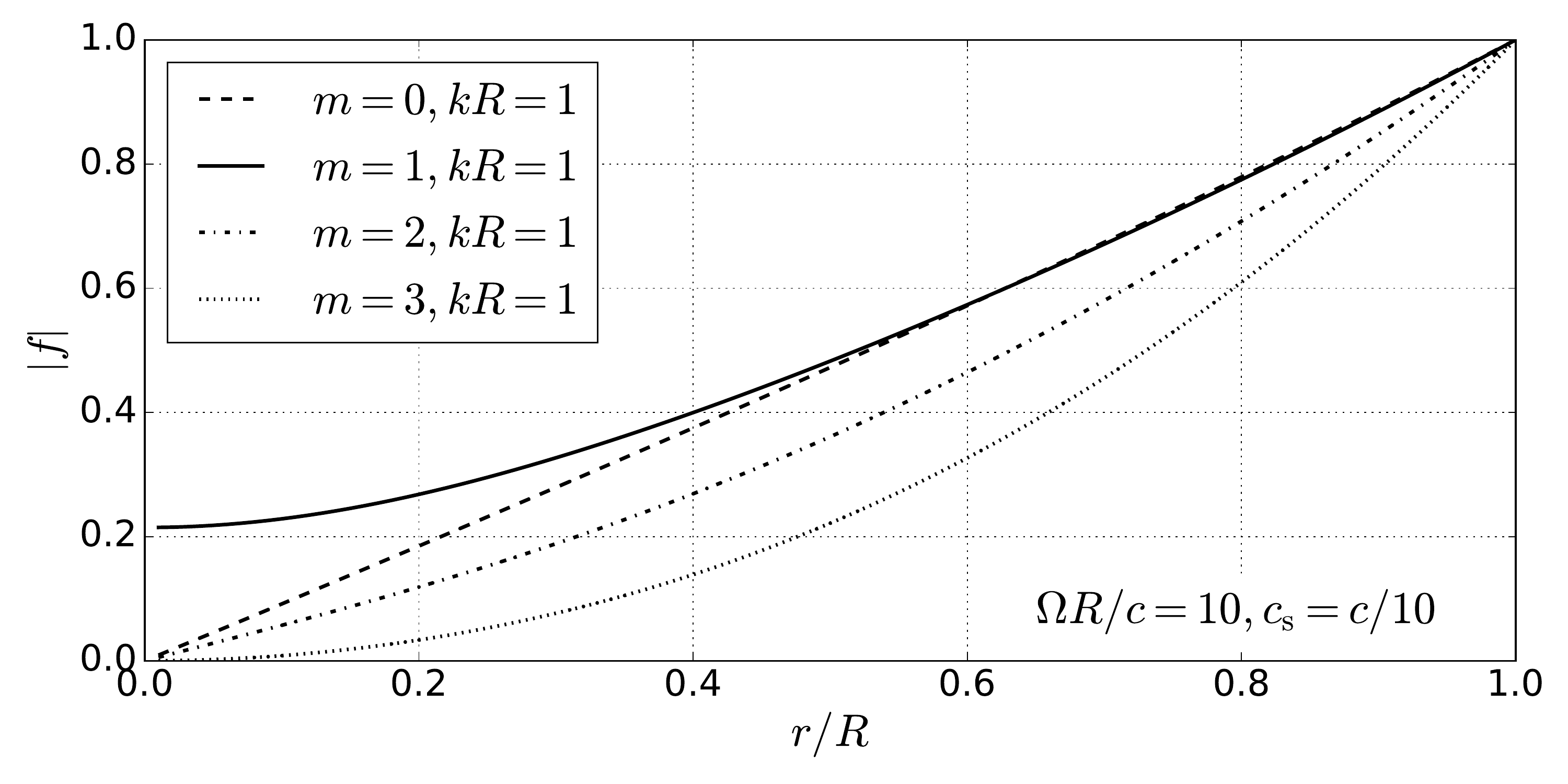}
\includegraphics[width=0.49\textwidth]{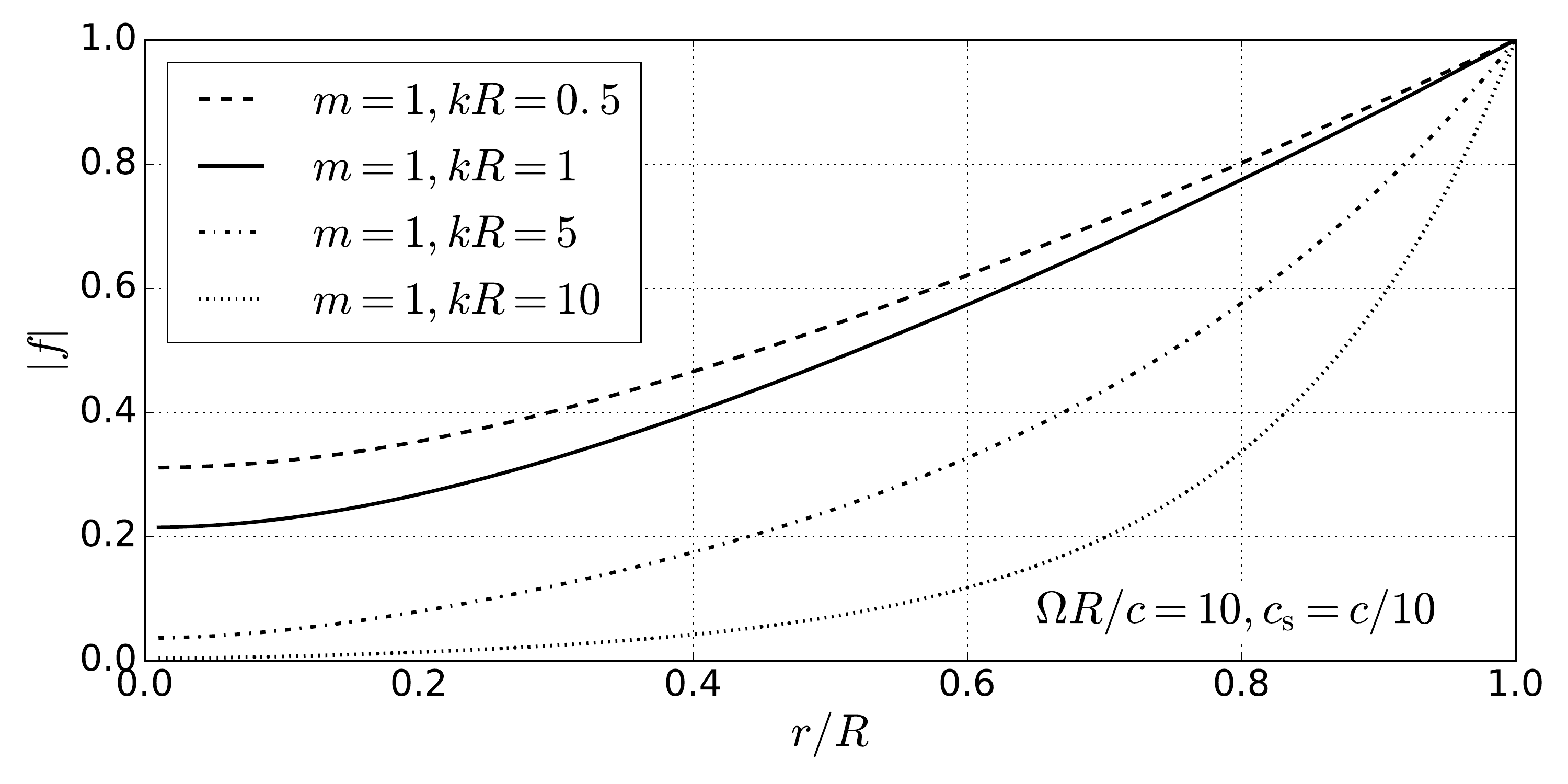}
\caption{Solution of Eq. \eqref{eq:main} for different values of $m$ and $kR$. We use $c_{\rm s}=c/10$ for the sound speed in the confining, unmagnetised medium and $\Omega R/c=10$ for the drift velocity of the magnetised plasma.}
\label{fig:f_cylindrical1}
\end{figure*}

\subsection{Pressure equilibrium at the jet's boundary}

\subsubsection{Perturbed magnetic pressure}

Using Eq. \eqref{eq:Etrans}-\eqref{eq:Btrans}, the magnetic pressure can be calculated as
\begin{equation}
\label{eq:pmag_complete}
p_{\rm mag}=\frac{\B^{*2}-\E^{*2}}{8\pi}=\frac{\psi'^2+I^2-\eta\phi'^2}{8\pi\eta}\;,
\end{equation}
where the prime denotes the derivative with respect to $\xi$. After expanding Eq. \eqref{eq:pmag_complete} to the first order in $f$, at the jet's boundary, which is defined by $\xi=R-f\left(R\right)$, one finds
\begin{equation}
\label{eq:pmag}
\delta p_{\rm mag} = \frac{1}{4\pi\eta}\left[\left(\eta\phi'\phi'' - \psi'\psi'' - II'\right)f+\left(\eta\phi'^2 - \psi'^2 \right)f'\right]\;,
\end{equation}
where $f'\equiv{\rm d}f/{\rm d}r$, and all the quantities are evaluated at $r=R$. Since we are not interested in perturbations at orders higher that first and Eq. \eqref{eq:pmag} is already linear in $f$, all the functions $\phi$, $\psi$, $I$ can be expressed in terms of the unperturbed fields; after straightforward calculations, inverting Eq. \eqref{eq:field_trans}-\eqref{eq:Btrans} one finds
\begin{align}
\phi' & = \frac{uB_\phi-E_{\rm r}}{\sqrt{1-u^2}} \\
\psi' & = kr\sqrt{1-u^2}B_{\rm z}-\frac{m\left(B_\phi-uE_{\rm r}\right)}{\sqrt{1-u^2}} \\
I & = mB_{\rm z}+kr\left(B_\phi-uE_{\rm r}\right) \;.
\end{align}

\subsubsection{Perturbed gas pressure}

Oscillations of the separation surface between the magnetised jet and the confining medium excite sound waves in the gas. The evolution of these waves is described by
\begin{equation}
\label{eq:sound}
\left(\frac{\partial^2}{\partial{\rm t}^2} - c_{\rm s}^2\nabla^2\right)\delta p_{\rm gas} = 0 \;,
\end{equation}
where $\delta p_{\rm gas}$ is the perturbation in the gas pressure. The speed of sound $c_{\rm s}$ is
\begin{equation}
\label{eq:cs}
c_{\rm s}^2=\frac{\Gamma p_{\rm gas}}{w_{\rm gas}}\;.
\end{equation}
For a cold gas ($p_{\rm gas}\ll\rho_{\rm gas}c^2$) the speed of sound reduces to the usual result $c_{\rm s}^2=\Gamma p_{\rm gas}/\rho_{\rm gas}$. In the opposite limit of a relativistically hot gas ($p_{\rm gas}\gg\rho_{\rm gas}c^2$) with $\Gamma=4/3$ we instead find $c_{\rm s}^2=\left(\Gamma-1\right)c^2=c^2/3$.

Since the problem has a cylindrical symmetry, the solution of Eq. \eqref{eq:sound} is of the form
\begin{equation}
\label{eq:deltarho}
\delta p_{\rm gas} = A {\rm e}^{i\left(kz+m\phi-\omega t\right)} H_{\rm m}^{\left(1\right)}\left(r\sqrt{\frac{\omega^2}{c_{\rm s}^2} - k^2}\right)\;,
\end{equation}
where $A$ is a scaling constant and $H_{\rm m}^{\left(1\right)}$ is the Hankel's function of order $m$ corresponding to an outgoing wave.

The scaling constant $A$ can be related to the amplitude of the oscillation of the last magnetic surface. Linearisation of the Euler's equation gives
\begin{equation}
\label{eq:euler_lin}
i\omega\delta v_{\rm r} = \frac{1}{w_{\rm gas}} \frac{\partial}{\partial{\rm r}}\left(\delta p_{\rm gas}\right) \;,
\end{equation}
where $\delta v_{\rm r}$ is the perturbation of the radial velocity. Since at first order the velocity of the last magnetic surface is purely radial, $\delta v_{\rm r}$ is related to $f$ by
\begin{equation}
\label{eq:vr}
\delta v_{\rm r} = \frac{\partial f}{\partial t} = -i\omega f\left(R\right)\;.
\end{equation}
Plugging Eq. \eqref{eq:deltarho} and \eqref{eq:vr} into Eq. \eqref{eq:euler_lin} one can express $A$ in terms of $f$. 
The perturbation to the gas pressure at $r=R$ is eventually
\begin{equation}
\label{eq:pgas}
\delta p_{\rm gas} = \frac{\omega^2 w_{\rm gas} H_{\rm m}^{\left(1\right)}\left(R\sqrt{\frac{\omega^2}{c_{\rm s}^2} - k^2}\right)}{\sqrt{\frac{\omega^2}{c_{\rm s}^2} - k^2}{H'}_{\rm m}^{\left(1\right)}\left(R\sqrt{\frac{\omega^2}{c_{\rm s}^2}-k^2}\right)} f \;,
\end{equation}
where ${H'}_{\rm m}^{\left(1\right)}$ denotes the derivative of the Hankel's function. Using Eq. \eqref{eq:balance} and \eqref{eq:cs}, the gas enthalpy can be finally calculated as $w_{\rm gas}=\Gamma p_{\rm mag}/c_{\rm s}^2$.

Hence, the condition of pressure equilibrium at the jet's boundary is finally
\begin{equation}
\label{eq:PB}
\delta p_{\rm mag} = \delta p_{\rm gas} \;,
\end{equation}
where $\delta p_{\rm mag}$ and $\delta p_{\rm gas}$ are given by Eq. \eqref{eq:pmag} and \eqref{eq:pgas} respectively. It is important to note that Eq. \eqref{eq:PB} depends on the speed of sound; hence, one would expect $c_{\rm s}$ to play an important role in determining the stability of the jet. For example, in the limit $c_{\rm s}\to 0$, one needs $\rho_{\rm gas}\to\infty$ to keep pressure equilibrium; hence, Eq. \eqref{eq:PB} reduces to $f\left(R\right)=0$, i.e. to a rigid wall placed at the jet's boundary.

\begin{figure}{\vspace{3mm}} 
\centering
\includegraphics[width=0.49\textwidth]{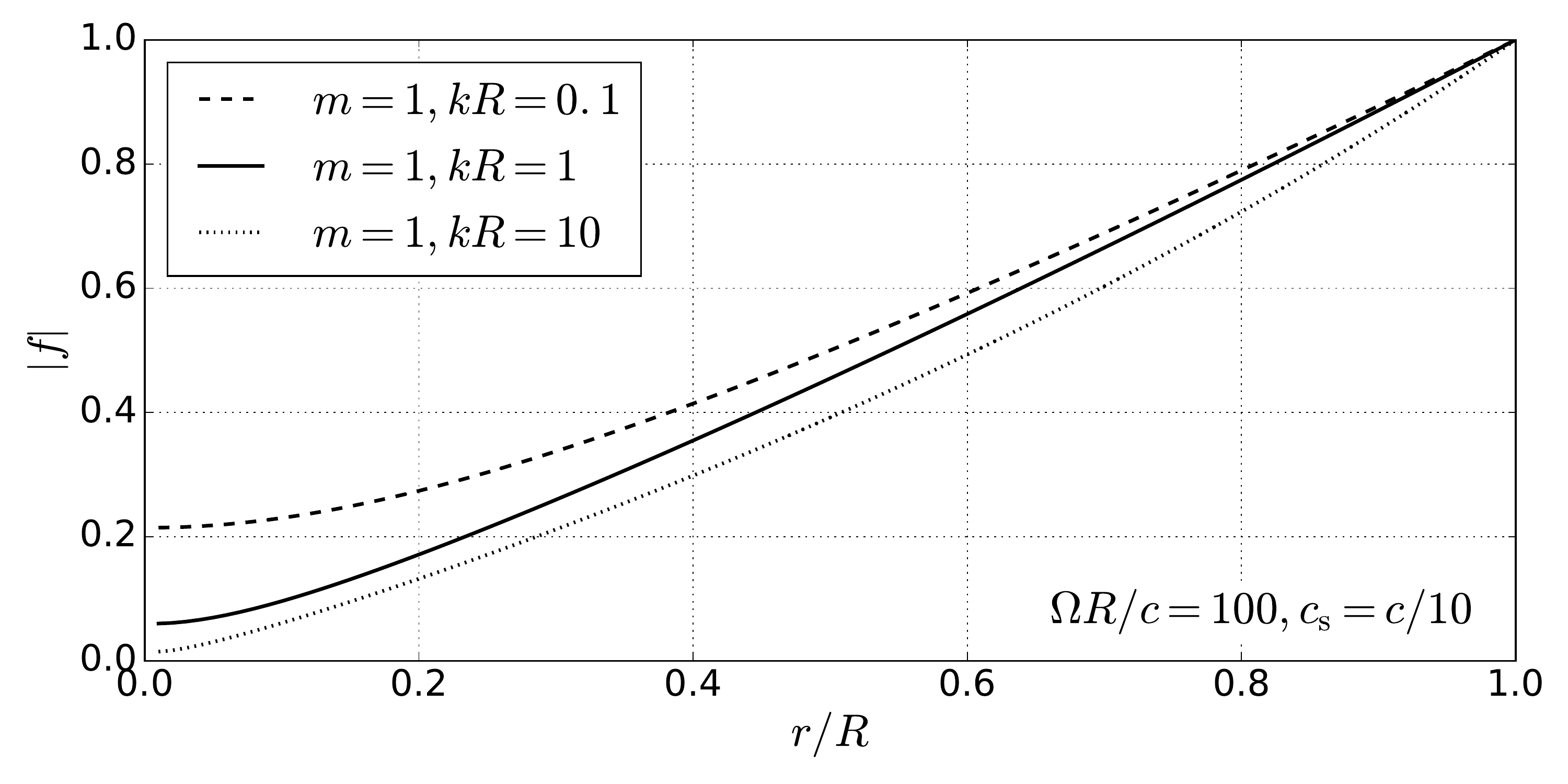}
\caption{Solution of Eq. \eqref{eq:main} for the $m=1$ mode and different values of $kR$. We use $c_{\rm s}=c/10$ and $\Omega R/c=100$.
}
\label{fig:f_cylindrical2}
\end{figure}

\section{Results}
\label{sec:results}

\subsection{Profile of the eigenmodes}

Jet's instabilities are a good candidate to convert the magnetic energy into the kinetic energy of the plasma. For this process to be efficient, the entire body of the jet needs to be involved. Hence, we are particularly interested in modes that manage to perturb a significant fraction of the jet, and are not restricted to the only surface.

In Figure \ref{fig:f_cylindrical1} we show the modulus of the solution of Eq. \eqref{eq:main} for different values of $m$ and $kR$, normalised to the value at $r=R$. We use $c_{\rm s}=c/10$ and $\Omega R/c=10$. The solution is generally concentrated at the jet's external boundary, as expected since these modes are excited by the interaction of the magnetised plasma with the confining gas. However, modes with a long wavelength are expected to be relevant over a significant range of radii: for example, the $m=1$, $kR=0.5$ mode has $\left|f\right|=0.5$ at $r/R=0.45$. As the wavelength decreases, the solution becomes more concentrated around $r\sim R$: in the left (right) panels we show such a trend for increasing values of $m$ ($k$).

In Figure \ref{fig:f_cylindrical2} we still take the same $c_{\rm s}=c/10$, but we consider a faster rotation, $\Omega R/c=100$. In this case the profile of the eigenmodes becomes less sensitive to the wavelength of the perturbation, and modes with shorter wavelengths can significantly perturb the jet's body. This is due to the fact that, when the velocity of the perturbations approaches the speed of light (which is easier in fast rotating jets), the effective wavelength in their own frame increases and perturbations become correspondingly less concentrated at the jet's boundary. Finally, note that $\left|f\right|$ is typically a factor $\gtrsim 5$ smaller at the centre than at the boundary of the jet; this may explain why the global structure of the jet is generally preserved in axisymmetric simulations (e.g. \citealt{KomissarovBarkov2007}), even if the surface is perturbed.

\subsection{Dispersion relation}
\label{sec:DR}

\begin{figure*}{\vspace{3mm}} 
\centering
\includegraphics[width=0.49\textwidth]{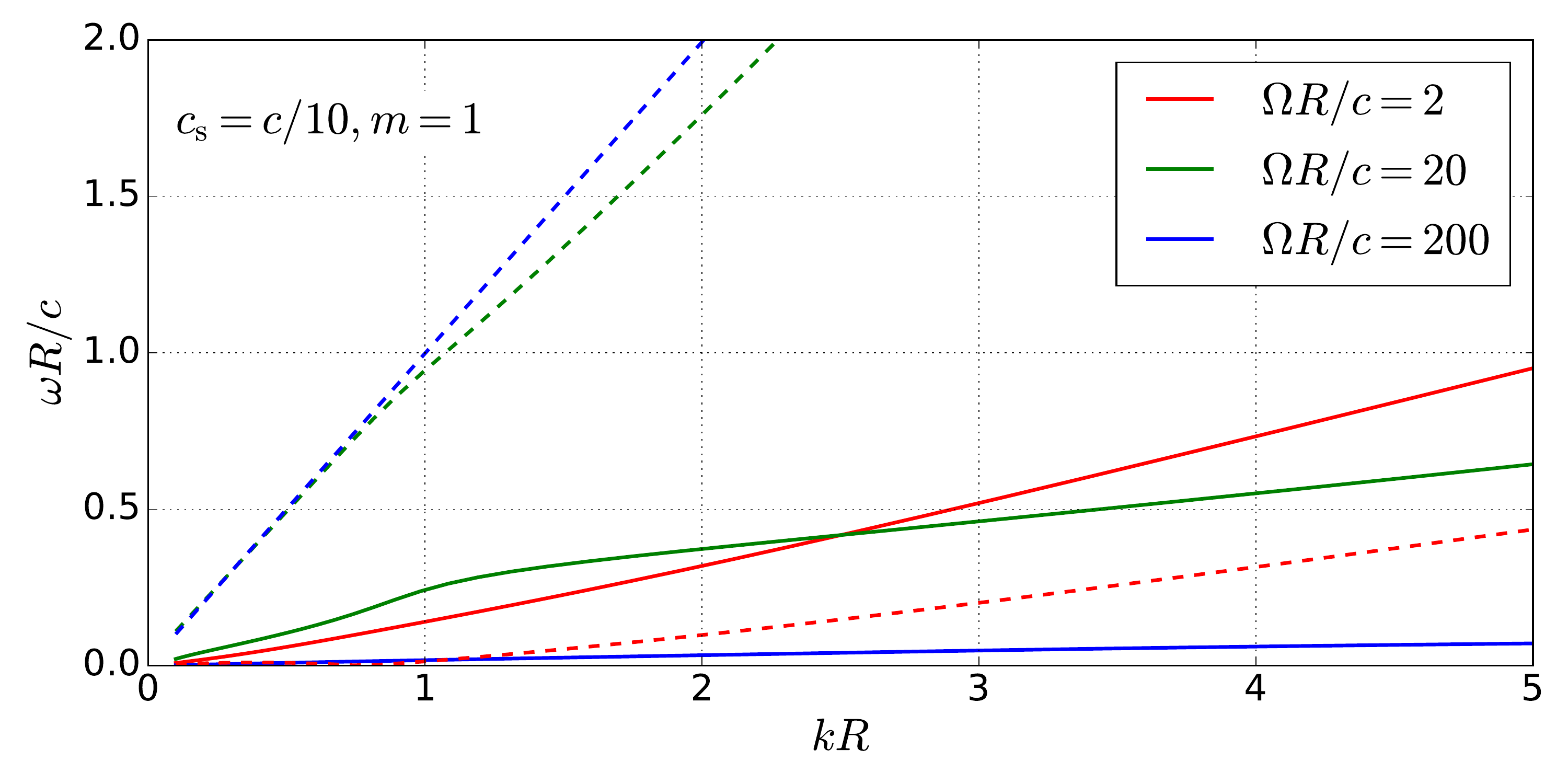}
\includegraphics[width=0.49\textwidth]{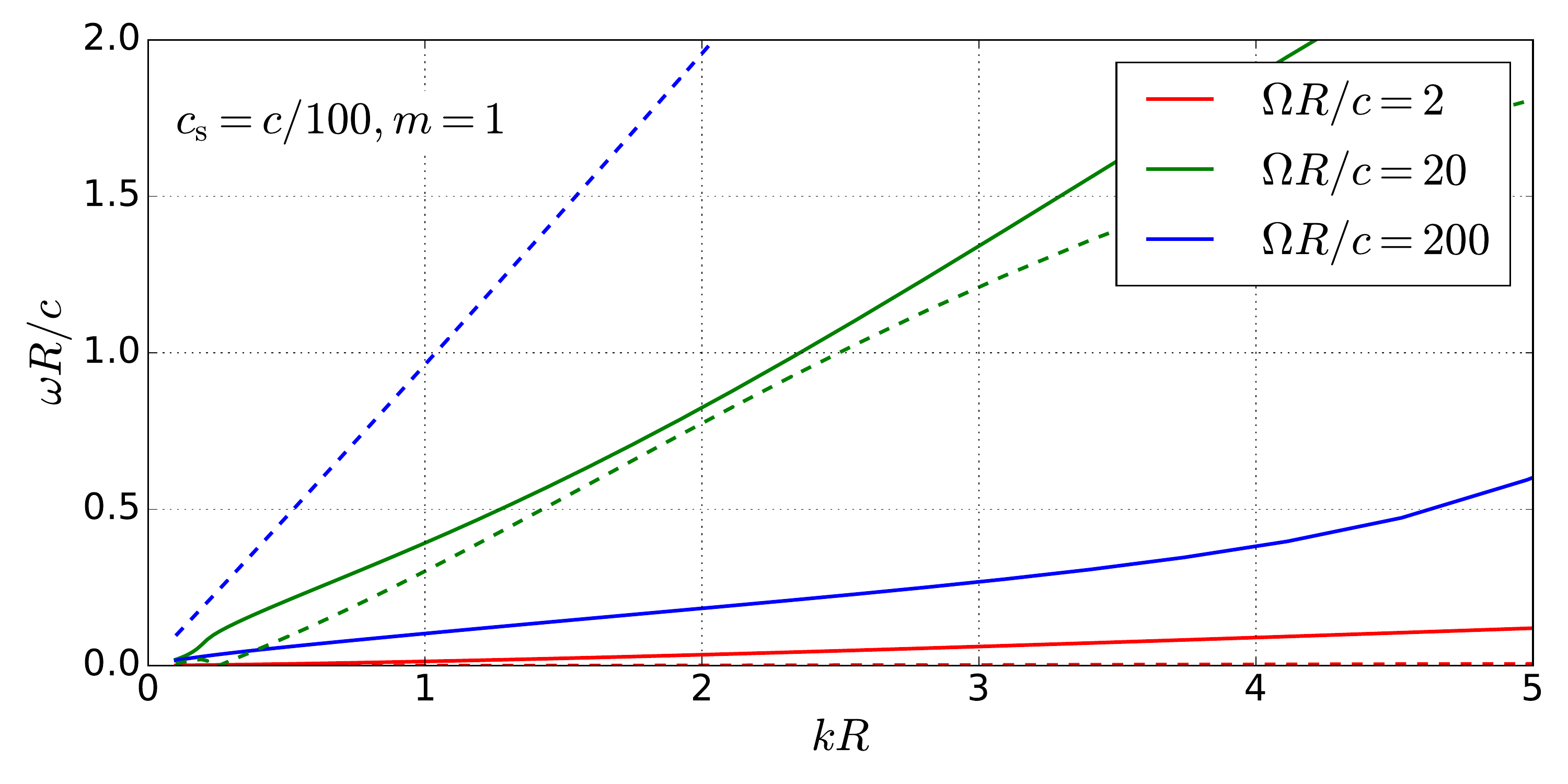}
\caption{Dispersion relation of the $m=1$ mode; solid (dashed) lines correspond to the imaginary (real) part of $\omega$. We use $c_{\rm s}=c/10$ ($c_{\rm s}=c/100$) for the sound speed in the confining medium in the left (right) panel.}
\label{fig:DR_cylindrical}
\end{figure*}

In Figure \ref{fig:DR_cylindrical} we show the dispersion relation of the $m=1$ mode,\footnote{Different values of $m$ (among the lowest ones,  which are the most relevant for the energy conversion) do not change significantly the dispersion relation. For example, when $m\leq 2$, $\omega$ changes by $\lesssim 50\%$ over the range $1\lesssim kR\lesssim 5$.} with the solid (dashed) lines corresponding to the imaginary (real) part of $\omega$. We use $c_{\rm s}=c/10$ ($c_{\rm s}=c/100$) in the left (right) panel and different values of $\Omega R/c$. The dispersion relation can be reasonably approximated by $\omega\propto k$, as expected for shear modes (note the analogy with the Kelvin-Helmholtz instability). Hence, modes with $kR\approx 1$ are the most relevant for energy conversion since they (i) grow faster than modes with smaller $k$; (ii) involve a significant fraction of the jet, unlike modes with higher $k$ which are more concentrated at the boundary.

The group velocity of the perturbation, $v_{\rm g}\equiv{\rm d}\omega_{\rm re}/{\rm d}k$, increases with $\Omega$, eventually approaching the speed of light in the limit $\Omega R/c\gg 1$. It is generally slower than the velocity of the magnetised plasma, but supersonic with respect to the confining gas; also note that $v_{\rm g}$ increases with $c_{\rm s}$. The growth rate of the perturbation, $\omega_{\rm im}$, shows a more complex behaviour: in the left panel of Figure \ref{fig:DR_cylindrical}, $\omega_{\rm im}$ remains almost constant while $\Omega R/c\leq 20$, while it is significantly lower when $\Omega R/c=200$; in the right panel, $\omega_{\rm im}$ is maximum when $\Omega R/c=20$, while it decreases for both $\Omega R/c=2$ and $\Omega R/c=200$. In Appendix \ref{sec:appendixA} we study the dispersion relation for a plane configuration of the electromagnetic fields, showing that it has a similar dependence on the parameters, and $\omega_{\rm im}$ indeed peaks when $\gamma\approx\Omega R/c\approx\sqrt{c/c_{\rm s}}$.

\subsection{Spatial growth of the perturbations}
\label{sec:spatial}

\begin{figure}{\vspace{3mm}} 
\centering
\includegraphics[width=0.49\textwidth]{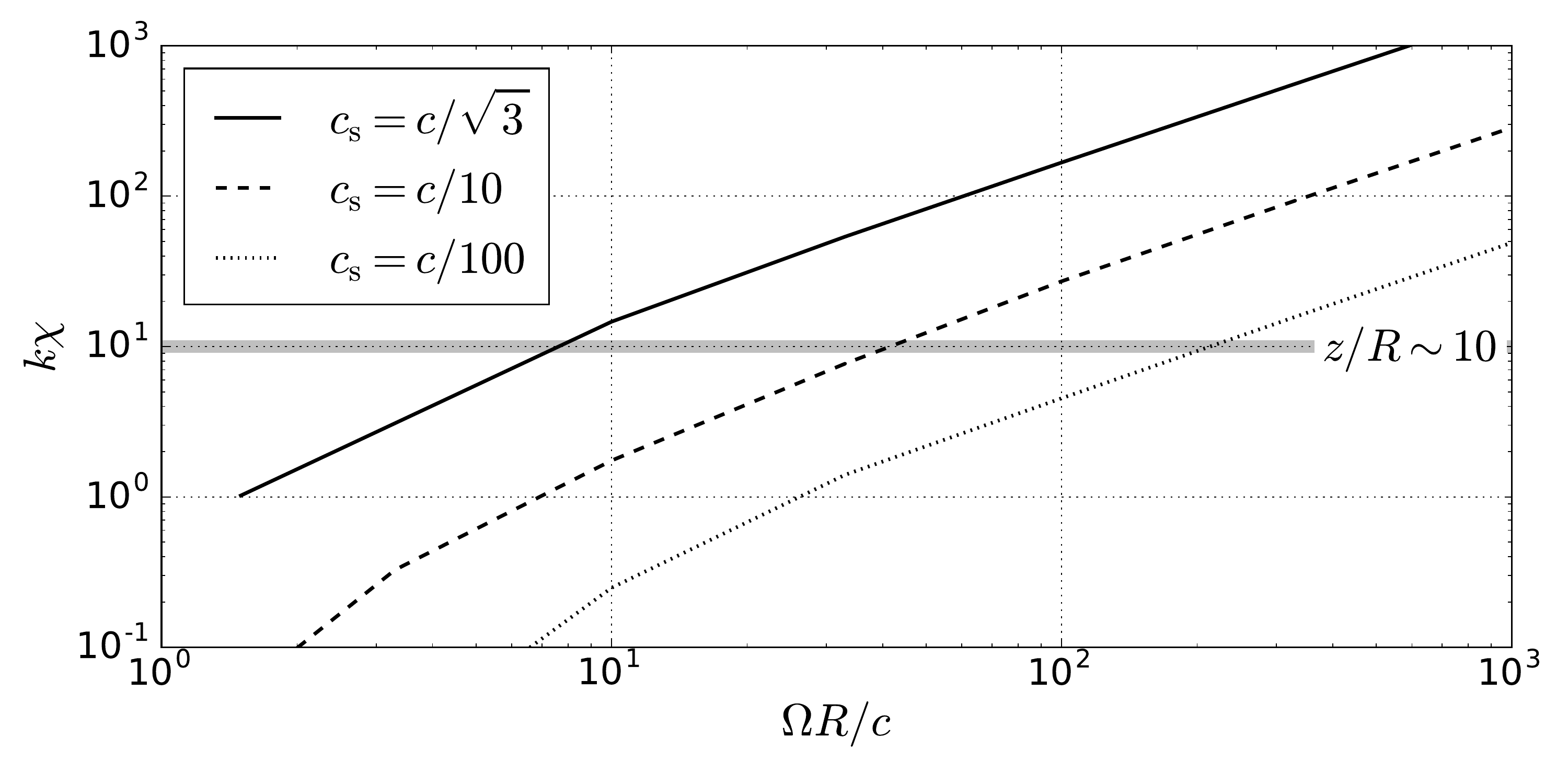}
\caption{Typical spatial scale $\chi$ over which the perturbations grow significantly, as a function of $\Omega R/c$. The grey, horizontal line shows our fiducial $z/R\sim 10$ (where $z$ is the typical distance from the source where most of the radiation is emitted).}
\label{fig:spatial}
\end{figure}

Since the perturbations move along the jet, it is important to identify the typical spatial scale $\chi$ over which they develop significantly. This is
\begin{equation}
\chi \approx \frac{v_{\rm g}}{\omega_{\rm im}}\approx\frac{\omega_{\rm re}}{\omega_{\rm im}}\frac{1}{k}\;,
\end{equation}
where we have approximated $v_{\rm g}\equiv{\rm d}\omega_{\rm re}/{\rm d}k\approx\omega_{\rm re}/k$; note that the ratio $\omega_{\rm re}/\omega_{\rm im}$ is almost independent on $k$. Surface modes strongly develop if $\chi\lesssim z$, where $z$ is the distance over which the jet doubles its radius. Due to the relativistic contraction of lengths in the frame of the perturbation, one expects modes with $kR\approx 1/\sqrt{1-v_{\rm g}^2/c^2}$ to perturb the jet significantly. Hence, the condition $\chi\lesssim z$ can be finally written as
\begin{equation}
\label{eq:condition}
k\chi\sqrt{1-\frac{v_{\rm g}^2}{c^2}}\lesssim\frac{z}{R}\;.
\end{equation}

In Figure \ref{fig:spatial} we show $k\chi$ as a function of $\Omega R/c$ for different values of $c_{\rm s}$.\footnote{Note that, due to the computational difficulties to precisely determine $v_{\rm g}$ when it approaches $c$, we are not considering the factor $\sqrt{1-v_{\rm g}^2/c^2}$. However, we expect this effect not to change our conclusions. This is discussed Appendix \ref{sec:appendixA}, where we study the simpler, plane configuration of the electromagnetic fields.} In the case of GRBs, we can approximate $z/R\sim 1/\theta_{\rm jet}$, where $\theta_{\rm jet}\sim 10\degree$ is the half-opening angle of the jet (e.g. \citealt{Frail2001, LeDermer2007}). In AGN, the radiation is typically emitted at a distance of hundreds/thousands of gravitational radii from the source (e.g. \citealt{Ghisellini2010}); the transverse scale of the jet in radiogalaxies, when resolved, is instead of the order of tens/hundreds of gravitational radii (e.g. \citealt{Mertens2016, Boccardi2016}). In both cases one finds $z/R\sim 10$, which is the fiducial value plotted with the grey, horizontal line in Figure \ref{fig:spatial}.

AGN jets, whose typical Lorentz factor is $\Omega R/c\lesssim 10$, manage to fulfil Eq. \eqref{eq:condition} even if the confining gas is relativistically hot. Hence, surface modes may contribute to dissipate the energy stored in the electromagnetic fields. On the other hand, GRBs are in the regime $\Omega R/c\gtrsim 100$, and their jet is confined by a relativistically hot cocoon. Hence, they violate Eq. \eqref{eq:condition} and the instability does not develop.

The impact of these modes on the jet's structure can be limited by non-linear effects saturating the instability. These are expected to be particularly relevant if (i) $\omega_{\rm im}\ll\omega_{\rm re}$, when the perturbations grow slowly with respect to the typical time scale of the system, or (ii) $c_{\rm s}\ll c$, in which case the velocity in the confining gas cannot be considered small. Finally note that, even if $\chi$ becomes arbitrarily small when $c_{\rm s}\ll c$, in this limit both $\omega_{\rm im}$ and $\omega_{\rm re}$ are vanishing, and the time for perturbations to develop diverges.

\subsection{Non-constant longitudinal magnetic field}
\label{sec:Bprofile}

\begin{figure}{\vspace{3mm}} 
\centering
\includegraphics[width=0.49\textwidth]{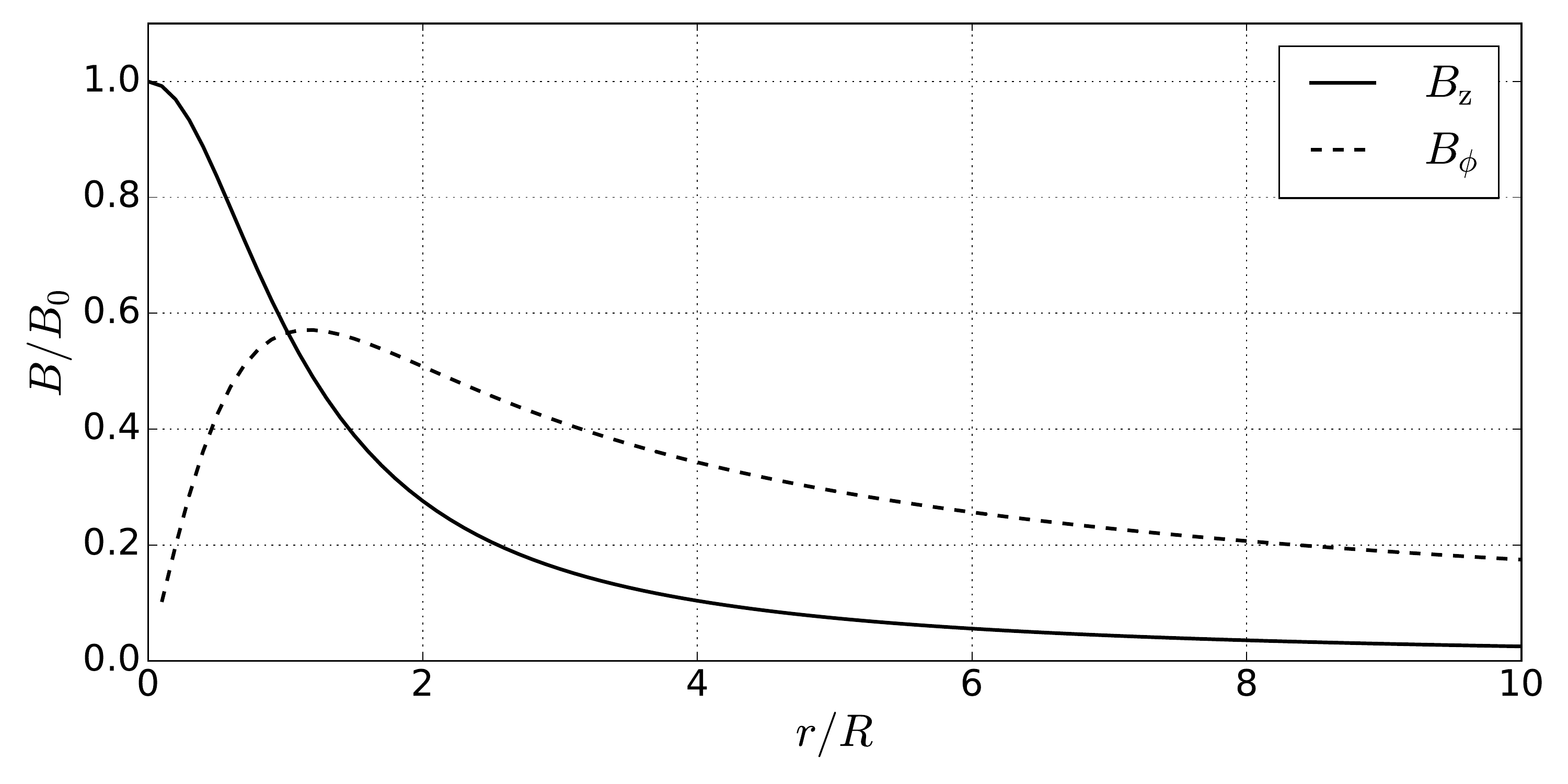}
\caption{Profile of the longitudinal (solid) and azimuthal (dashed) magnetic fields used in Section \ref{sec:Bprofile}.}
\label{fig:Bprofile}
\end{figure}

In this section we are using a more general solution of Eq. \eqref{eq:equilibrium}, allowing for a dependence of $B_{\rm z}$ on $r$. Specifically, we take
\begin{equation}
\label{eq:Bsol}
B_{\rm z} = \frac{B_0}{\left[1+\left(r/R\right)^2\right]^\alpha}\;,
\end{equation}
where we use a fiducial $\alpha=0.8$. Plugging this $B_{\rm z}$ profile into Eq. \eqref{eq:equilibrium}, one gets (see for example \citealt{Mizuno2012})
\begin{equation}
\label{eq:solution}
B_\phi=-\left[P^2 + \left(\frac{\Omega r}{c}\right)^2\right]^{1/2} B_{\rm z}\;,
\end{equation}
where
\begin{equation}
\label{eq:Psol}
P^2 = \frac{\left(r_0/r\right)^2\left[1+\left(r/r_0\right)^2\right]^{2\alpha}-\left(r_0/r\right)^2-2\alpha}{2\alpha-1}\;.
\end{equation}
In Figure \ref{fig:Bprofile}, the solid (dashed) line shows the radial profile of $B_{\rm z}$ ($B_\phi$), where we are using $\Omega R/c=1$. Note that most of the jet's energy is concentrated at $r\lesssim R$; an additional length scale needs to be introduced, i.e. the distance where the magnetic pressure is balanced by the confining medium. In the following we denote this scale as $L$, and we study the cases $L\gg R$ and $L\sim R$.

\begin{figure*}{\vspace{3mm}} 
\centering
\includegraphics[width=0.49\textwidth]{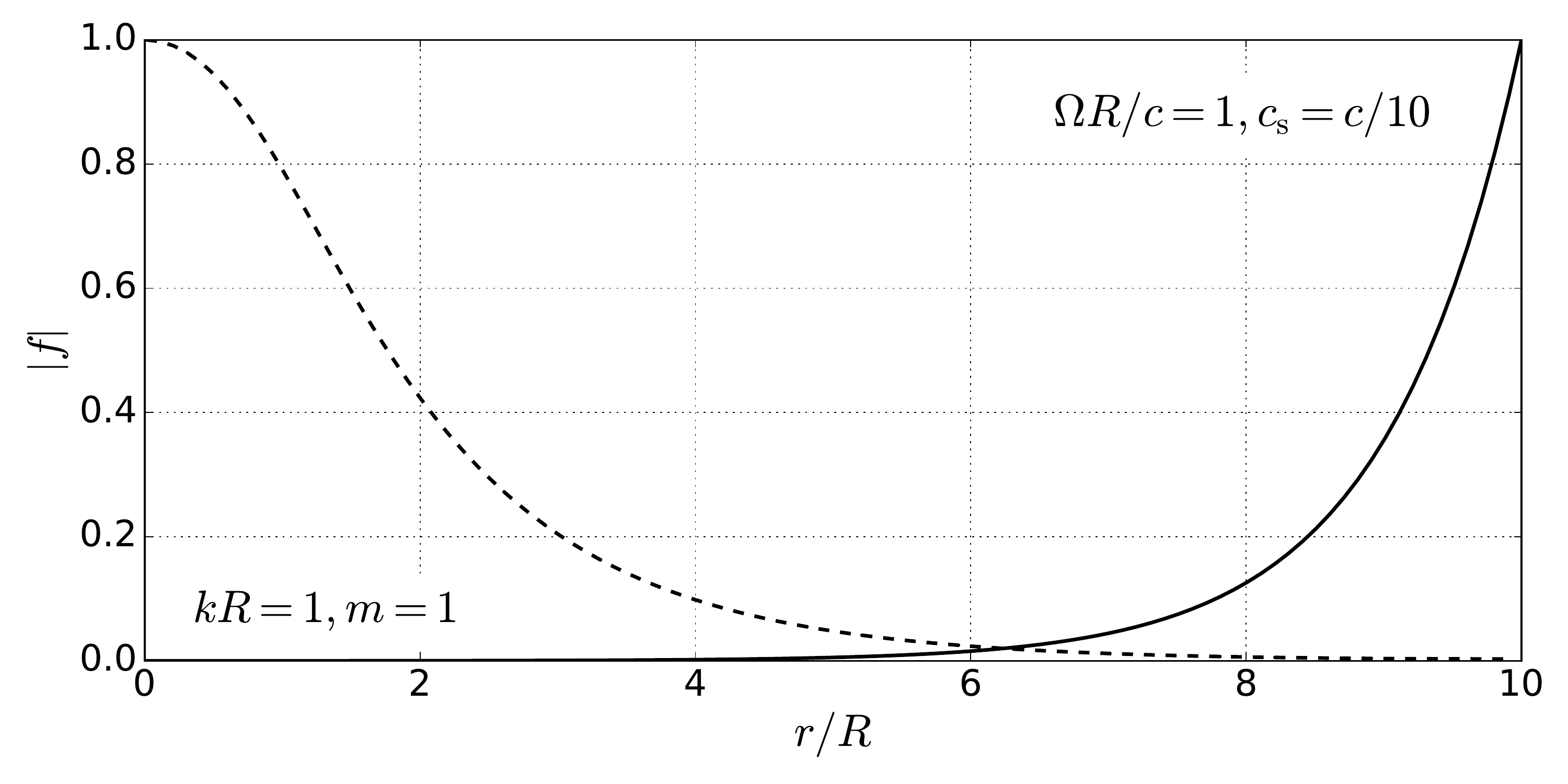}
\includegraphics[width=0.49\textwidth]{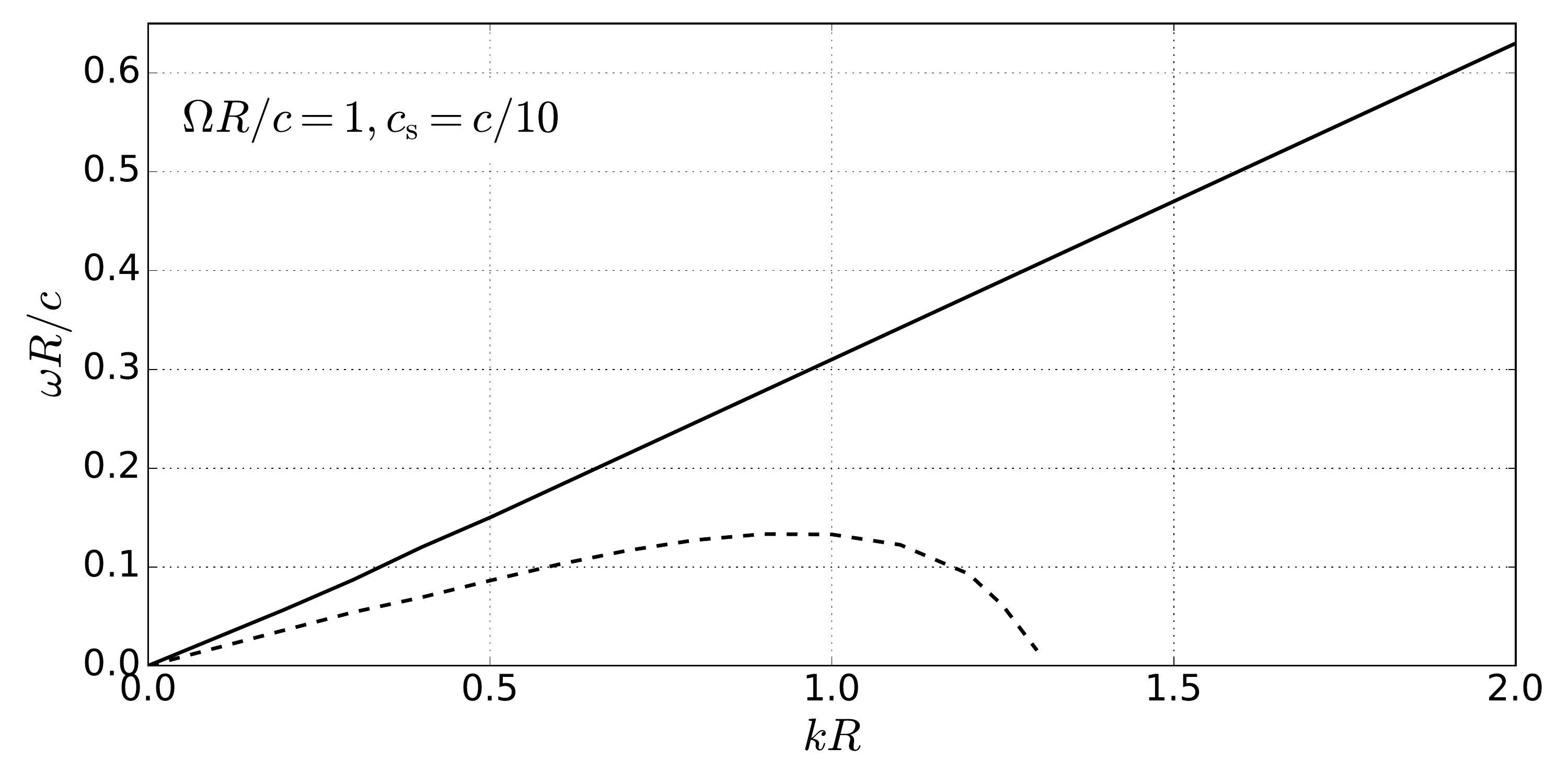}
\caption{The case of a non-constant longitudinal magnetic field (we use $\alpha=0.8$). We show the solution of Eq. \eqref{eq:main} for the $m=1$, $kR=1$ mode (left panel) and the corresponding dispersion relations (right panel). We use $c_{\rm s}=c/10$ for the sound speed in the confining, unmagnetised medium and $\Omega R/c=1$ for the drift velocity of the magnetised plasma. The magnetised jet extends to $L=10R$; at this scale the magnetic pressure is balanced by the pressure of the confining, unmagnetised gas.}
\label{fig:DR_modes}
\end{figure*}

When $B_{\rm z}$ is not constant, the jet becomes unstable also for perturbations peaked at the jet's core (e.g. \citealt{Lyubarski1999, Sobacchi2017}). The most dangerous perturbation corresponds to the $m=1$ mode (e.g. \citealt{Bateman1978}). In the left panel of Figure \ref{fig:DR_modes} we show the solution of Eq. \eqref{eq:main} for the electromagnetic field configuration described above, using $L/R=10$. There are two different solution: the first one (solid line) is peaked at the boundary of the jet, and closely resembles the surface modes that we have discussed; the second one (dashed line) is peaked at the core, and corresponds to the eigenfunctions of the modes shown in \citet{Sobacchi2017}. This is confirmed by the corresponding dispersion relation (right panel), where we show the imaginary part of $\omega$. The solid line is well approximated by $\omega\propto k$, as the surface modes analysed above; the dashed line is instead peaked at $kR\sim 1$, in agreement with the results of \citet{Sobacchi2017}.

\begin{figure*}{\vspace{3mm}} 
\centering
\includegraphics[width=0.49\textwidth]{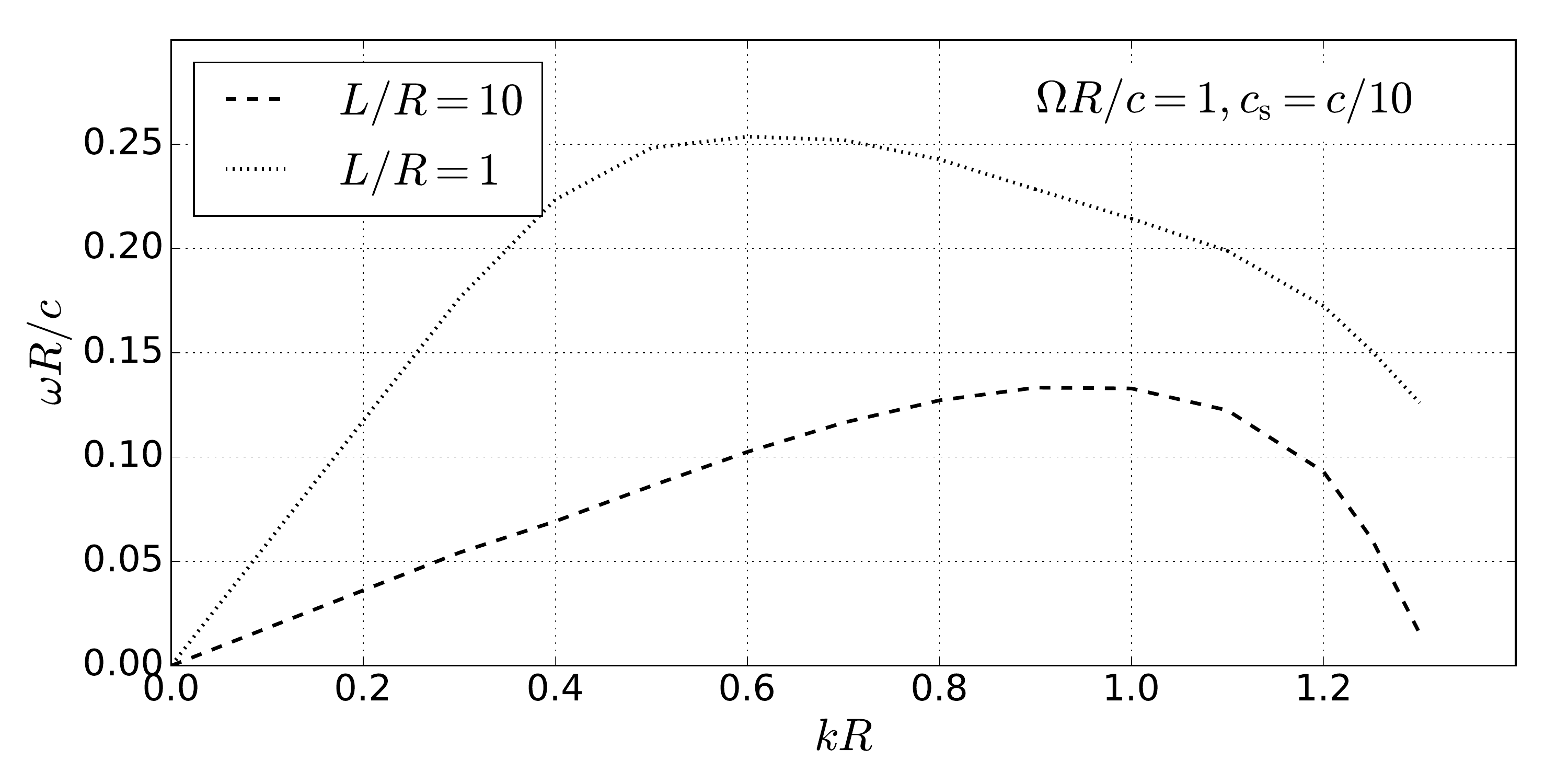}
\includegraphics[width=0.49\textwidth]{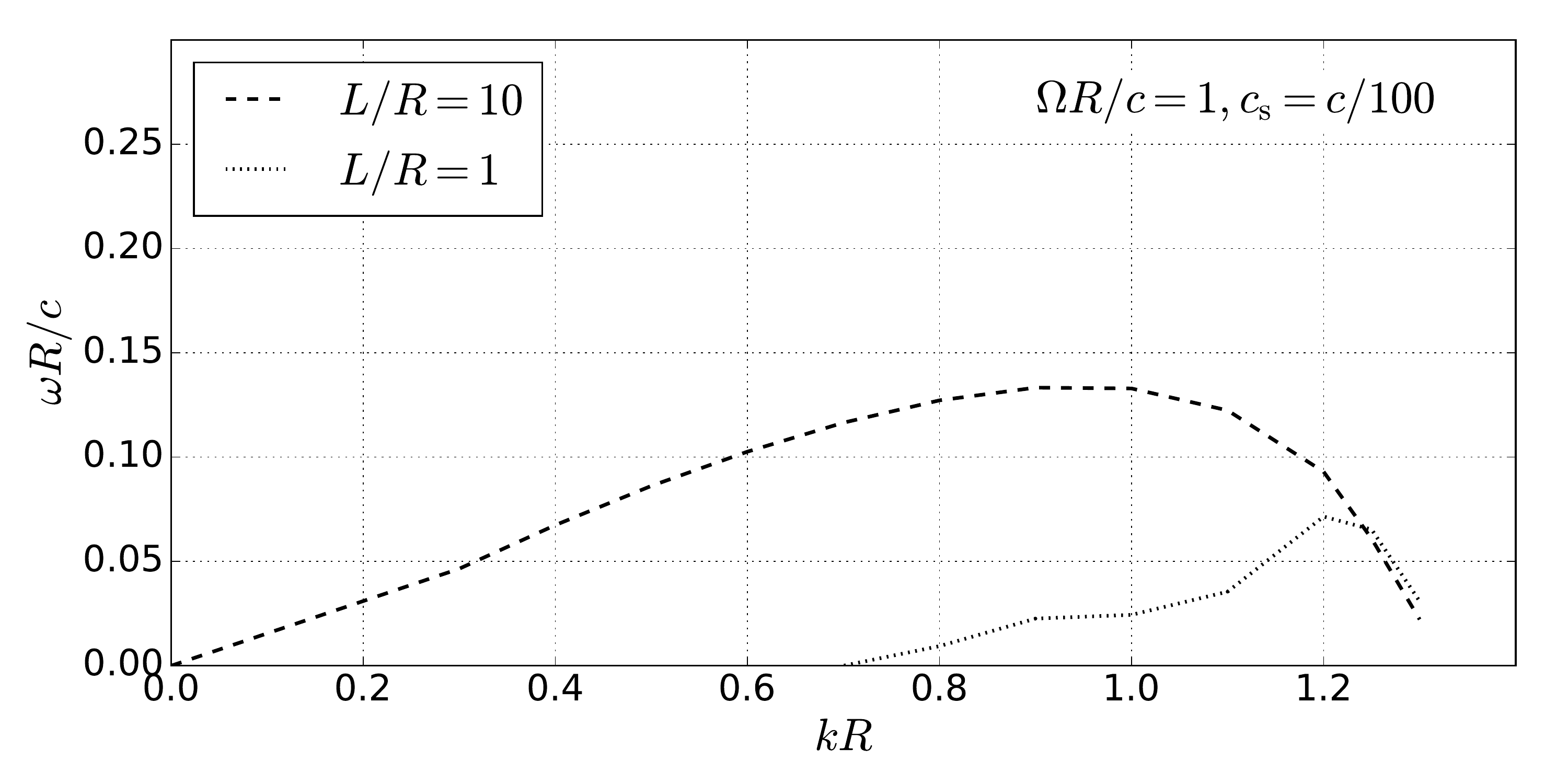}
\caption{Dispersion relation of the core modes with a non-constant longitudinal magnetic field (we use $\alpha=0.8$). The left (right) panels correspond to $c_{\rm s}=c/10$ ($c_{\rm s}=c/100$). The dashed (dotted) lines in each panel correspond to the case where the magnetised jet extends to $L=10R$ ($L=R$).}
\label{fig:DR_modes_cs}
\end{figure*}

In Figure \ref{fig:DR_modes_cs} we study the dependence of the dispersion relation of the core modes on the scale $L$ where the transition from the magnetised jet to the confining medium occurs. In the left panel we use a relatively high $c_{\rm s}=c/10$: surprisingly, we find that the lower $L/R=1$ corresponds to a faster growth rate of the instability. However, this trend is reversed for lower $c_{\rm s}$: in the right panel we show the case $c_{\rm s}=c/100$, where the instability is severely suppressed when $L/R=1$. Also note that, since the core modes are not sensitive to the presence of a boundary placed at $L\gg R$, the dispersion relation do not change from the left to the right panel when $L/R=10$.

\section{Conclusions}
\label{sec:conclusions}

In this paper we have studied the stability of relativistic, force-free jets, focusing on the impact of the external confining medium. We have considered an idealised, cylindrical configuration with a sharp transition between the magnetised jet and the gas (assumed to be at rest) which confines the jet through its pressure. Our main results can be summarised as follows:
\begin{enumerate}
\item the contact surface is unstable for shear modes, which are excited by the velocity gradient between the jet and the confining medium. When the wavelength is comparable to the size of the jet ($kR\lesssim 1$), the perturbation affects a significant fraction of the jet's body;
\item the dispersion relation of these surface modes is analogous to the Kelvin-Helmholtz one, i.e. $\omega\propto k$. The most relevant modes (which grow fast and perturb the jet significantly) have $kR\approx 1$;
\item when the Lorentz factor of the magnetised plasma is $\Gamma\sim10$ ($\Gamma\sim 100$), these perturbations typically develop after propagating along the jet for tens (hundreds) of jet's radii. Surface modes may therefore play an important role in converting the energy of the jet from the Poynting flux to the kinetic energy of the plasma, particularly in AGN;
\item the growth rate of the surface modes peaks when $\Omega R/c~\approx~\sqrt{c/c_{\rm s}}$, where $\Omega$ is the angular velocity of the field lines and $c_{\rm s}$ the sound speed of the confining gas.
\end{enumerate}
The contribution of these surface modes is expected to be particularly relevant if the longitudinal magnetic field is approximately constant, since in this case the jet is stable to core modes.

\section*{Acknowledgements}

The authors acknowledge support from the Israeli Science Foundation under Grant No. 719/14.

\def\aap{A\&A}\def\aj{AJ}\def\apj{ApJ}\def\apjl{ApJ}\def\mnras{MNRAS}
\def\araa{ARA\&A}\def\physrep{PhR}\def\sovast{Sov. Astron.}\def\nar{NewAR}
\def\aapr{Astronomy \& Astrophysics Review}\def\apjs{ApJS}\def\nat{Nature}\def\na{New Astron.}
\def\prd{Phys. Rev. D}\def\pre{Phys. Rev. E}\def\pasp{PASP}
\bibliographystyle{mn2e}
\bibliography{2d}

\appendix

\section{The plane configuration}
\label{sec:appendixA}

\begin{figure*}{\vspace{3mm}} 
\centering
\includegraphics[width=0.49\textwidth]{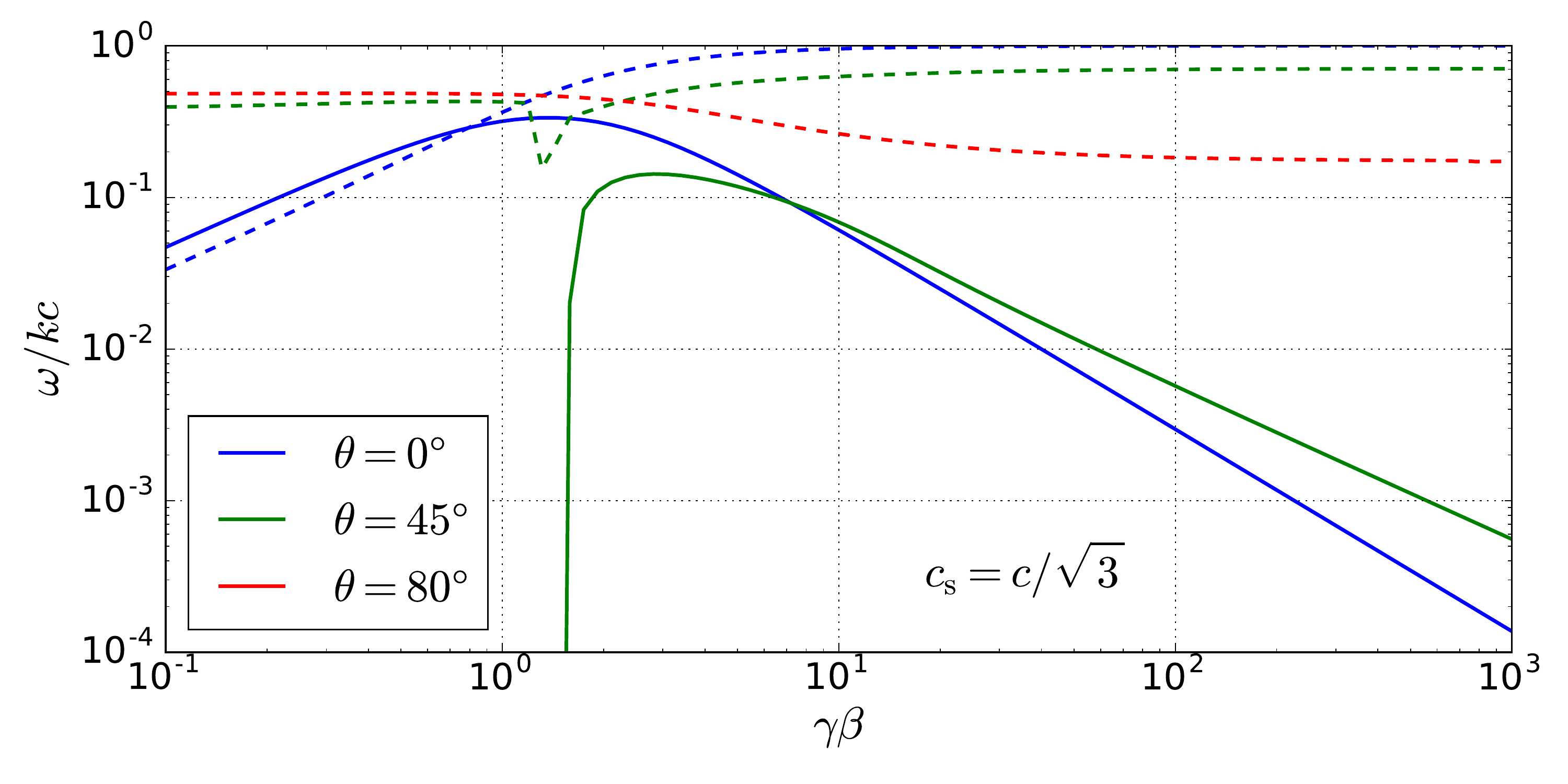}
\includegraphics[width=0.49\textwidth]{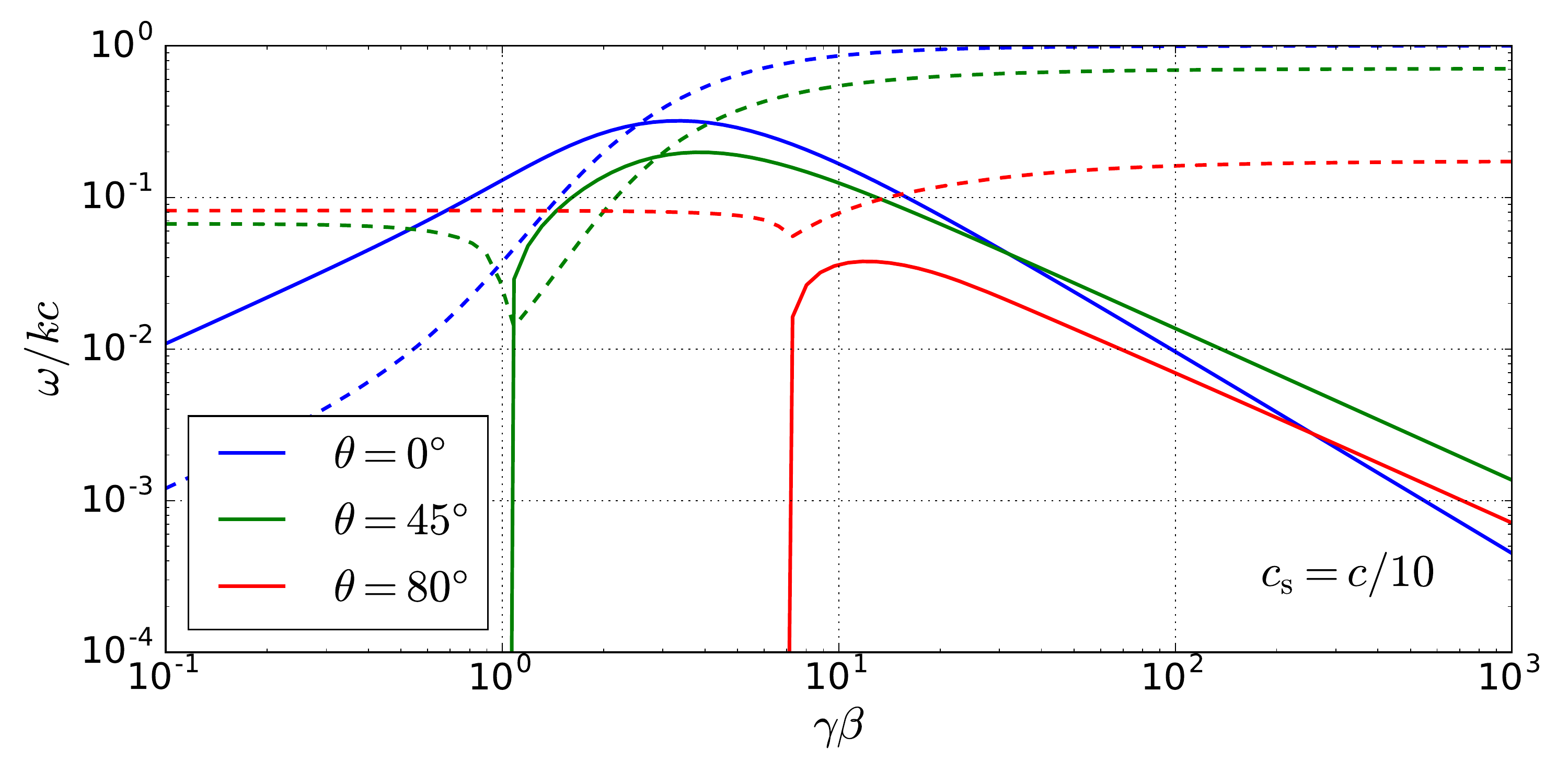}
\includegraphics[width=0.49\textwidth]{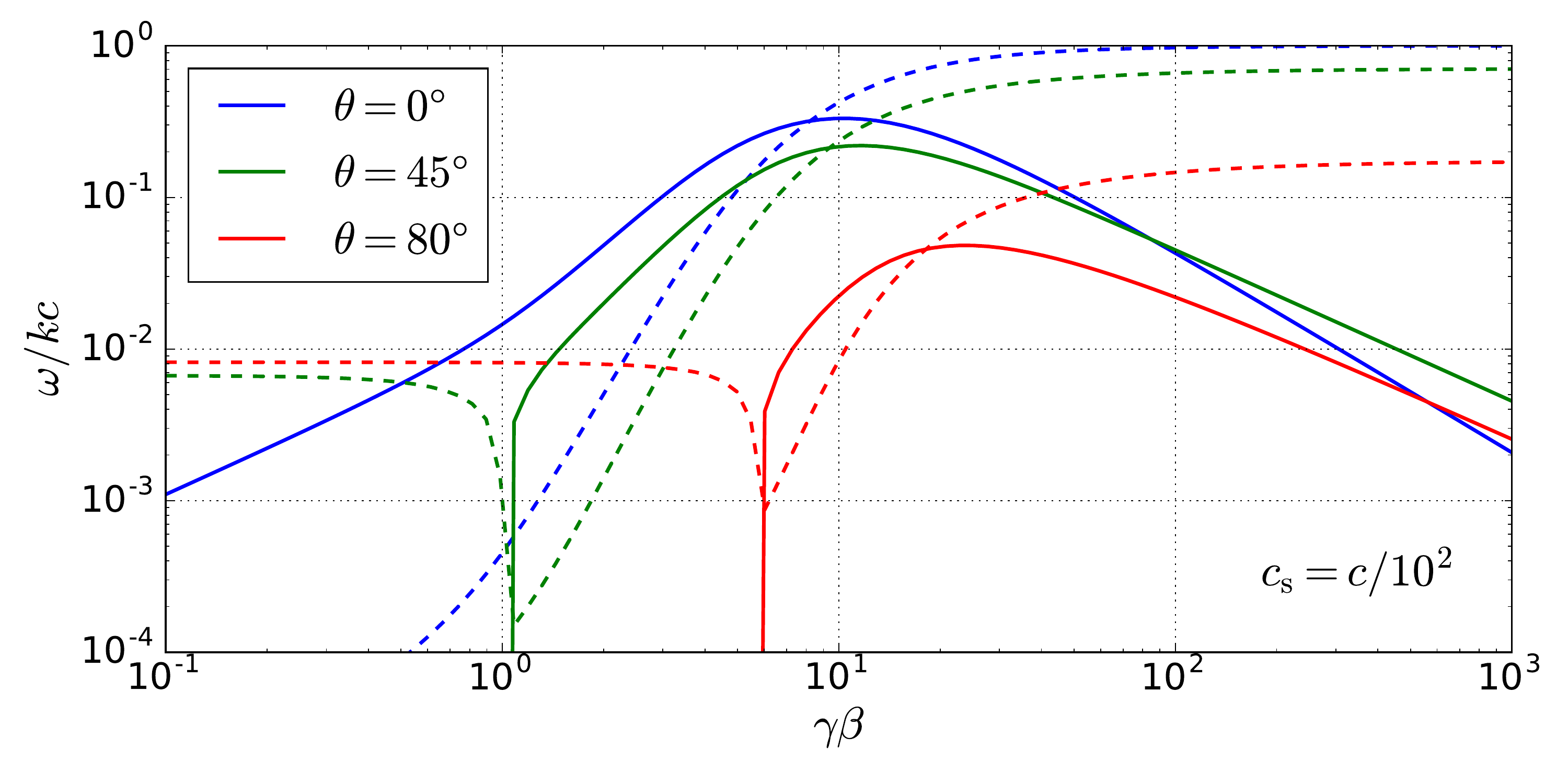}
\includegraphics[width=0.49\textwidth]{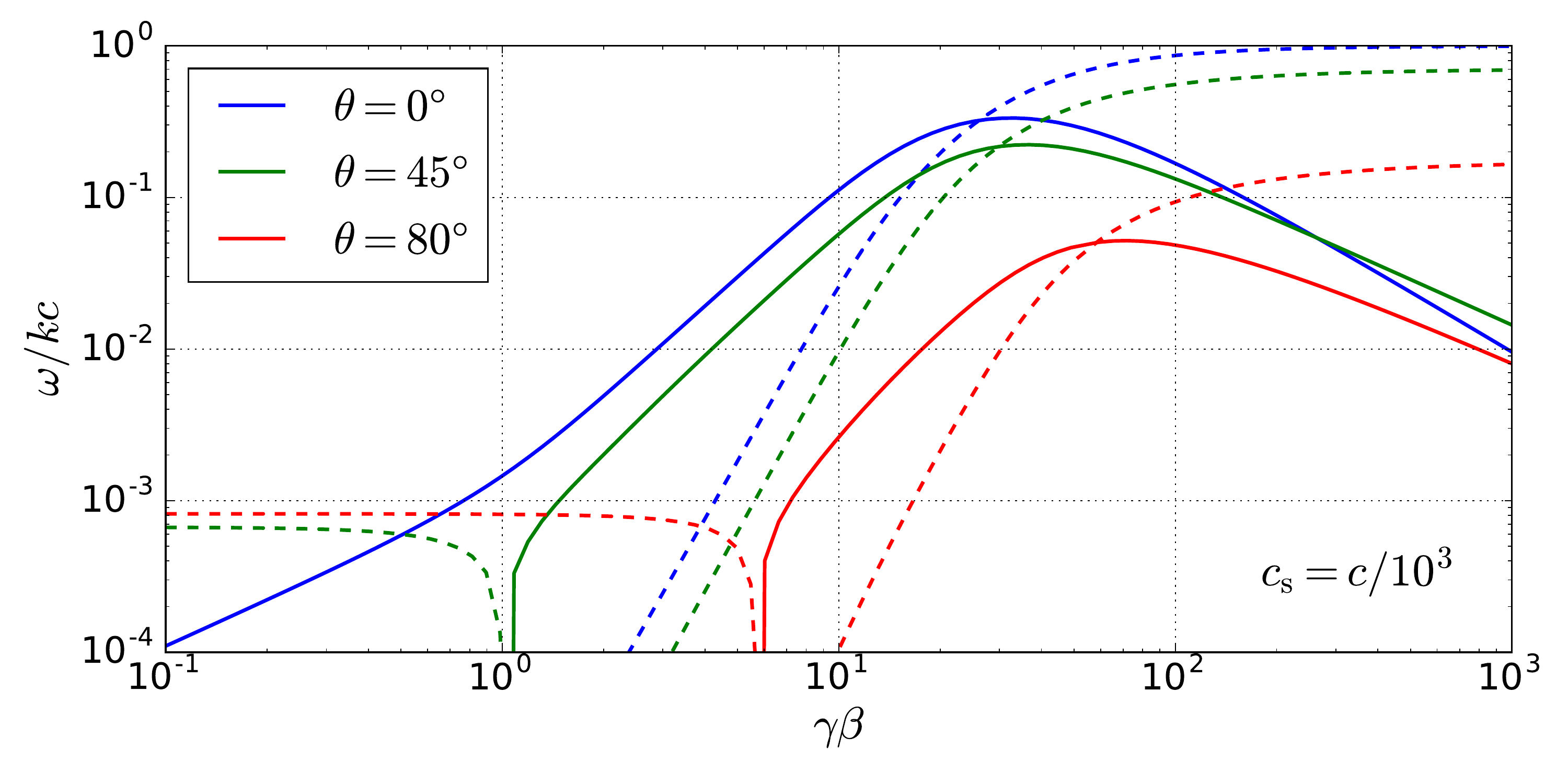}
\includegraphics[width=0.49\textwidth]{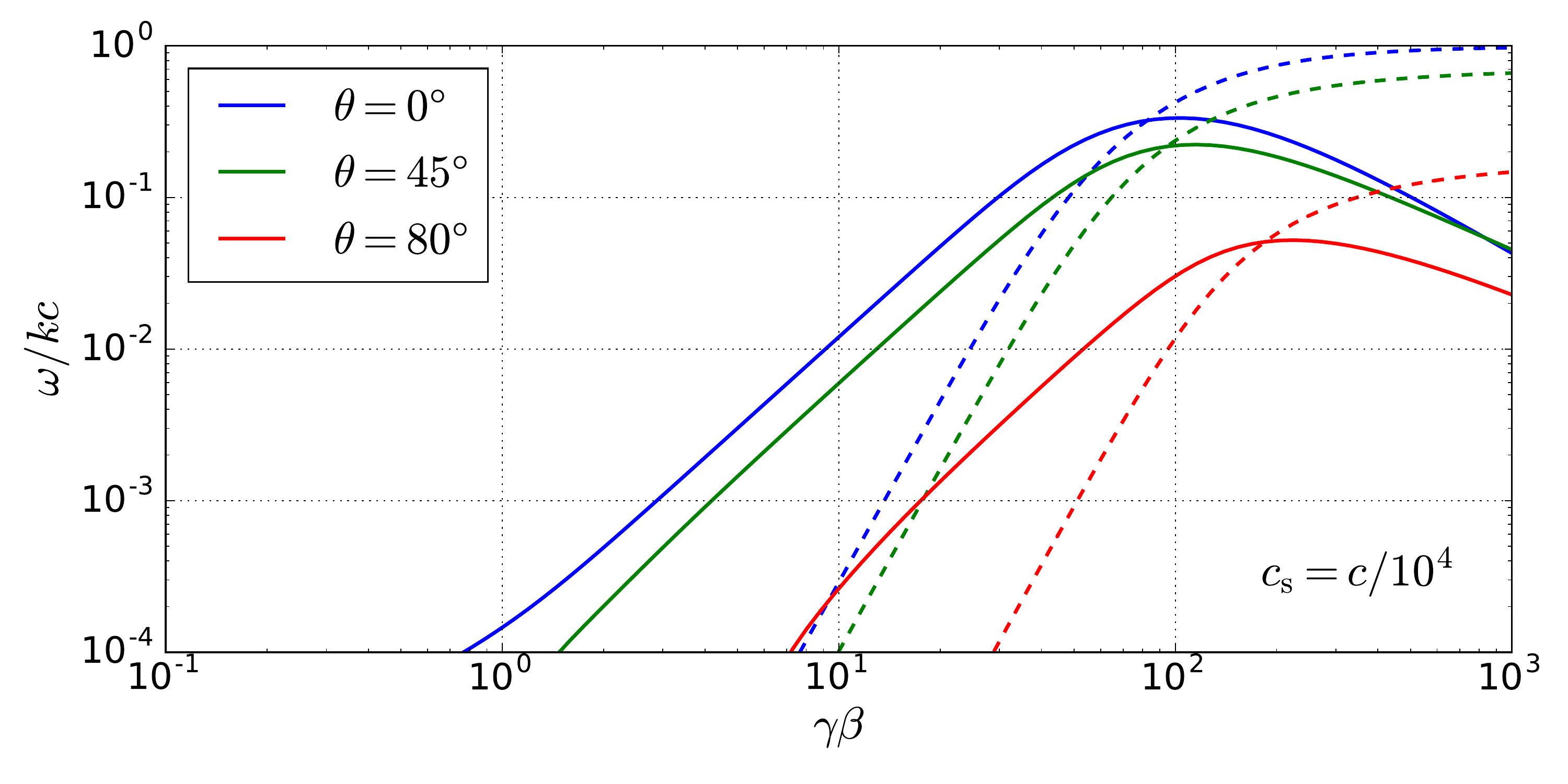}
\includegraphics[width=0.49\textwidth]{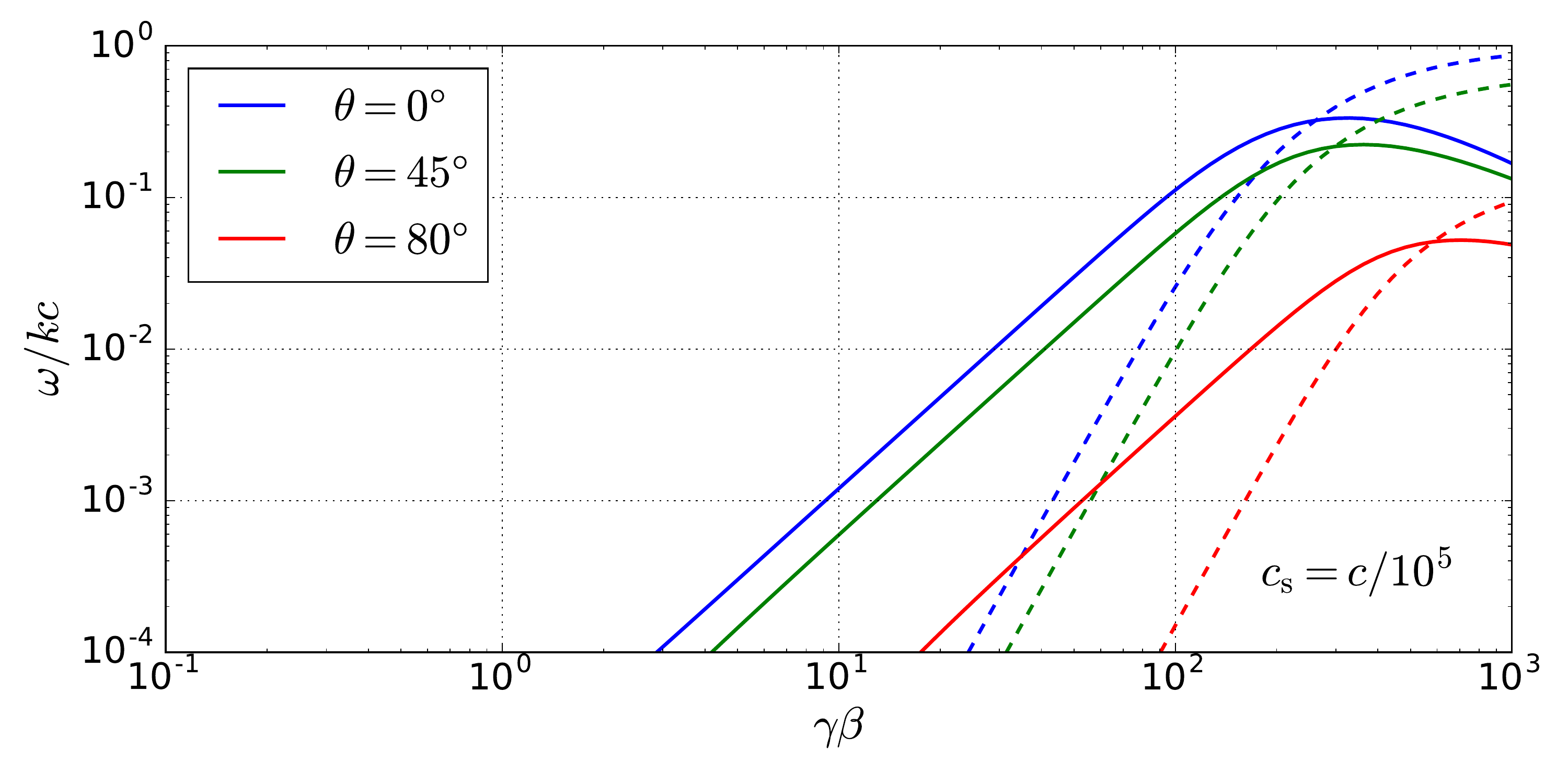}
\caption{Dependence of the proper frequency $\omega$ on the drift velocity $\gamma\beta$ of the magnetised plasma (solid/dashed lines correspond to the imaginary/real part of $\omega$ respectively). We show different values of (i) the sound speed $c_{\rm s}$ in the confining, unmagnetised medium and (ii) the angle $\theta$ between $k$ and the drift velocity. }
\label{fig:DR_plane}
\end{figure*}

We consider the same problem in a plane configuration, with constant electromagnetic fields in the half-plane $x<0$. In this case, the transverse displacement of the magnetic surfaces is simply described by
\begin{equation}
\label{eq:f}
f\propto\exp\left(kx\sqrt{1-u^2}\right)\;.
\end{equation}
We put the $z$ axis along the wavelength of the perturbation, so that we can take $m=0$, and we study Eq. \eqref{eq:PB} in the limit $kR\to\infty$.\footnote{Of course, Eq. \eqref{eq:DR_plane} can be also derived directly, and not as a limit of Eq. \eqref{eq:PB}. Here we do not provide such a derivation.} The dispersion relation can be finally written as
\begin{align}
\label{eq:DR_plane}
2\left[\sin^2\theta - \frac{\left(\beta-u\cos\theta\right)^2}{1-u^2}\right] & \left(1-u^2\right)^{1/2} \left(1-\frac{c^2}{c_{\rm s}^2}u^2\right)^{1/2} = \nonumber\\
& = \Gamma\left(1-\beta^2\right)\frac{c^2}{c_{\rm s}^2}u^2\;,
\end{align}
where $\beta=E/B$ is the drift velocity of the magnetised plasma and $\theta$ is the angle between $\pmb{\beta}\equiv\E\times\B/B^2$ and ${\bf k}$. Note that Eq. \eqref{eq:DR_plane} depends on $\omega$ and $k$ only through the combination $u\equiv\omega/kc$; hence, in the plane configuration $\omega$ is exactly proportional to $k$.

\begin{figure}{\vspace{3mm}} 
\centering
\includegraphics[width=0.49\textwidth]{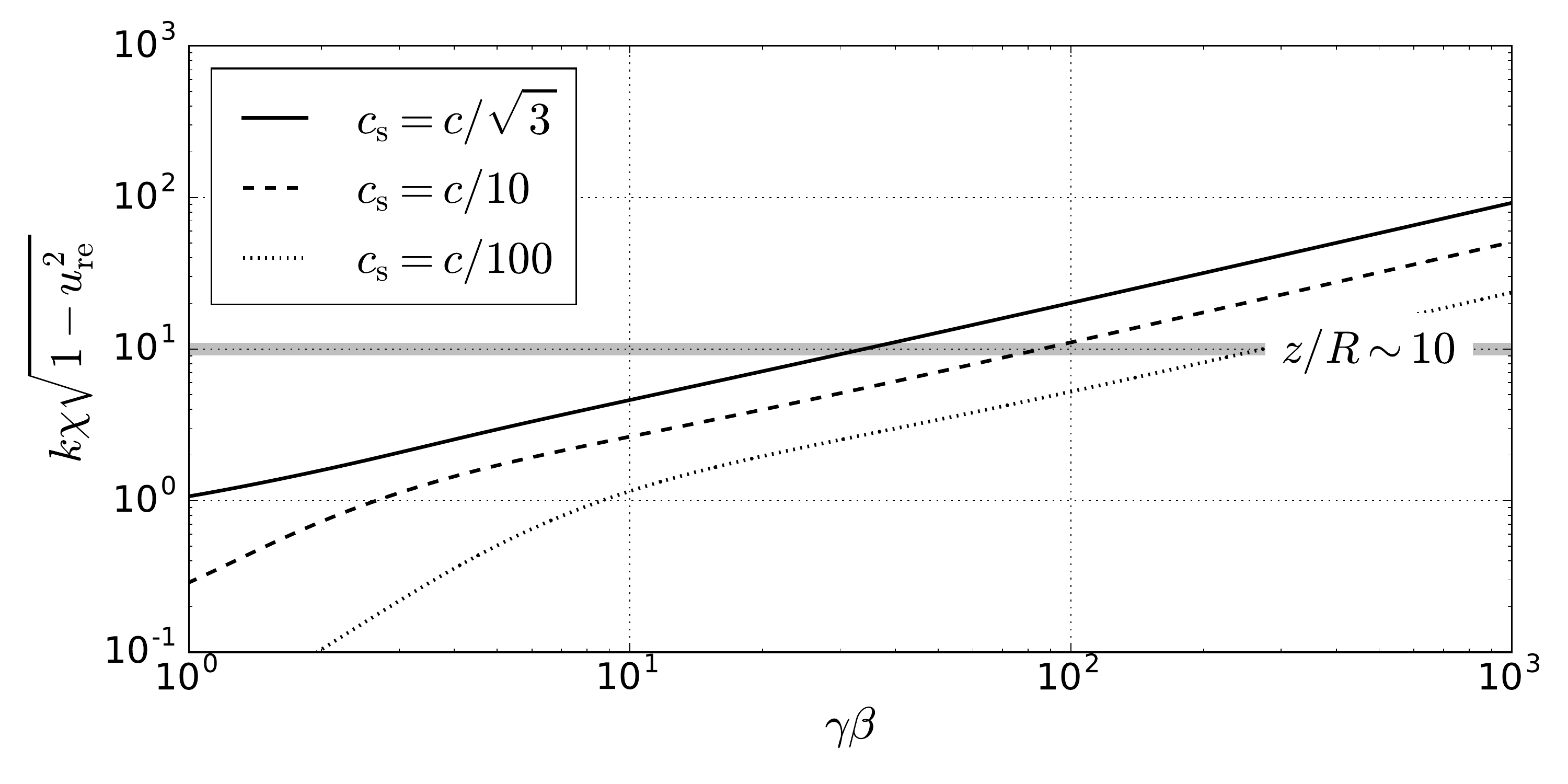}
\caption{Same as Figure \ref{fig:spatial}, but considering the effect of relativistic length contraction. We show the most unstable modes, corresponding to $\theta=0\degree$.}
\label{fig:spatial2}
\end{figure}

In Figure \ref{fig:DR_plane} we show the solution of Eq. \eqref{eq:DR_plane} for different values of $\theta$ and of the drift velocity of the magnetised plasma, $\gamma\beta$. Solid (dashed) lines correspond to the imaginary (real) part of $\omega$. The largest values of $\theta$ are more stable, and instability generally disappears when $\beta\lesssim\sin\theta$. This is due to the face that, for $\theta~=~90\degree$, the wavelength of the perturbation is exactly parallel to the magnetic field, and the magnetic tension stabilises the perturbations.

The panels correspond to different values of $c_{\rm s}$, from $c_{\rm s}~=~c/\sqrt{3}$ (top-left) to $c_{\rm s}=c/10^5$ (bottom-right). The unstable modes can be classified according to three regimes: (i) when $\beta\ll 1$ the proper frequency scales as $\omega_{\rm re}\propto\beta$ and $\omega_{\rm im}\propto\beta$, in agreement with the usual, non-relativistic Kelvin-Helmholtz dispersion relation; (ii) when $1\ll\gamma\ll\sqrt{c/c_{\rm s}}$ we have $\omega_{\rm re}\propto\gamma^3$ and $\omega_{\rm im}\propto\gamma^2$; (iii) when $\gamma\gg\sqrt{c/c_{\rm s}}$, the drift velocity saturates and $\omega_{\rm re}$ does not depend on $\gamma$, while the growth rate is relativistically suppressed, $\omega_{\rm im}\propto 1/\gamma$. In the third regime the frame moving along $\pmb{\beta}$ with the same phase velocity of perturbations indeed reaches a relativistic speed, and the corresponding Lorentz factor scales as $\gamma\sqrt{c_{\rm s}/c}$. Finally, note that the growth rate of the perturbations peaks when $\gamma\beta\approx\sqrt{c/c_{\rm s}}$, and the maximum is $\omega_{\rm im}/kc\approx 0.3$.

It is useful to reconsider in this simplified context the spatial scale $\chi$ over which the perturbations grow. In Figure \ref{fig:spatial2} we show $k\chi\sqrt{1-u_{\rm re}^2}$ as a function of $\gamma\beta$ (note that $u_{\rm re}\equiv v_{\rm g}/c$). A comparison with Figure \ref{fig:spatial}, where we did not include the effect of relativistic length contraction, shows that our main conclusions remain unchanged.

\subsubsection*{Physical interpretation}

In order to gain some physical insight on the dispersion relation, it is useful to rewrite Eq. \eqref{eq:DR_plane} for $\theta=0\degree$, when the magnetic field has not any stabilising effect on the perturbations. Note that $\theta=~0\degree$ indeed provides a reasonable description of the most unstable modes. We find
\begin{equation}
\label{eq:DR_simple}
\frac{2}{\Gamma}\frac{\left(\beta-u\right)^2}{\sqrt{1-u^2}} + \frac{c^2}{\gamma^2c_{\rm s}^2} \frac{u^2}{\sqrt{1-u^2c^2/c_{\rm s}^2}} = 0\;,
\end{equation}
where $\gamma=1/\sqrt{1-\beta^2}$. Since pressure equilibrium requires $B^2/\gamma^2\sim w_{\rm gas}c_{\rm s}^2$, the ratio between the energy density of the confining gas and that stored in the electromagnetic fields is approximately $c^2/\gamma^2c_{\rm s}^2$.

We can now make a comparison with the usual setup of the Kelvin-Helmholtz instability. Consider two compressible fluids of densities $\rho_1$, $\rho_2$ having sound speeds $c_1$, $c_2$ respectively. Assume that a sharp transition occurs between the first fluid moving with velocity $V$ and the second fluid at rest (note that pressure equilibrium requires $\rho_1 c_1^2 = \rho_2 c_2^2$). Finally, neglect the effect of gravity, and take all the velocities to be non-relativistic. Following \citet{Landau}, one can write the dispersion relation as
\begin{equation}
\label{eq:KH}
\frac{\rho_1\left(V-v\right)^2}{\sqrt{1-\left(V-v\right)^2/c_1^2}} + \frac{\rho_2v^2}{\sqrt{1-v^2/c_2^2}} = 0\;,
\end{equation}
where $v\equiv\omega/k$ is the phase velocity of the perturbation, assumed to be parallel to $V$. Apart from numerical factors of order unity, Eq. \eqref{eq:DR_simple} and \eqref{eq:KH} presents several analogies: (i) the terms $\left(\beta-u\right)^2$, $u^2$ and $\left(V-v\right)^2$, $v^2$ respectively; (ii) the terms $c^2/\gamma^2c_{\rm s}^2$ and $\rho_2/\rho_1$; (iii) the square roots at the denominator, which depend on the compressibility of the fluids and approach unity if $V$ is subsonic.

Eq. \eqref{eq:KH} is well known to describe stable modes when $c_1=~c_2<V/\sqrt{8}$; in this case the instability appears only when $v$ is not aligned with $V$ (e.g. \citealt{Landau}). However, since sound waves propagate at the speed of light in the force-free plasma, the closest case to that studied in the paper would be $c_2 < V < c_1$. We tested that in this regime Eq. \eqref{eq:KH} indeed predicts perturbations to be unstable.

\end{document}